\documentclass[12pt]{article}
\usepackage{epsfig,amsmath,amsfonts,amssymb,amstext,afterpage,psfrag,slashbox}
\long\def\symbolfootnote[#1]#2{\begingroup%
\def\thefootnote{\fnsymbol{footnote}}\footnote[#1]{#2}\endgroup}

\def\l{\langle}
\def\r{\rangle}

\def\spose#1{\hbox to 0pt{#1\hss}}
\def\lsim{\mathrel{\spose{\lower 3pt\hbox{$\mathchar"218$}}
 \raise 2.0pt\hbox{$\mathchar"13C$}}}
\def\gsim{\mathrel{\spose{\lower 3pt\hbox{$\mathchar"218$}}
 \raise 2.0pt\hbox{$\mathchar"13E$}}}

\catcode`@=11
\def\@citex[#1]#2{%
  \if@filesw\immediate\write\@auxout{\string\citation{#2}}\fi
  \def\@citea{}\@cite{\@for\@citeb:=#2\do
    {\@citea\def\@citea{,\penalty\@m}\@ifundefined
      {b@\@citeb}{{\bf ?}\@warning
{Citation `\@citeb' on page \thepage \space undefined}}%
      \hbox{\csname b@\@citeb\endcsname}}}{#1}}
\def\citer{\@ifnextchar [{\@tempswatrue\@citexr}{\@tempswafalse\@citexr[]}}
  \def\@citexr[#1]#2{%
    \if@filesw\immediate\write\@auxout{\string\citation{#2}}\fi
    \def\@citea{}\@cite{\@for\@citeb:=#2\do
      {\@citea\def\@citea{--\penalty\@m}\@ifundefined
{b@\@citeb}{{\bf ?}\@warning
{Citation `\@citeb' on page \thepage \space undefined}}%
\hbox{\csname b@\@citeb\endcsname}}}{#1}}

\begin{document}

\begin{titlepage}

\begin{flushright}
{\small
LMU-ASC~19/12\\ 
FLAVOUR(267104)-ERC-9\\
March 2012\\
}
\end{flushright}

\vspace{0.5cm}
\begin{center}
{\Large\bf \boldmath                                               
Effective Theory of a Dynamically Broken\\      
\vspace*{0.3cm}                                                            
Electroweak Standard Model at NLO                     
\unboldmath}
\end{center}

\vspace{0.5cm}
\begin{center}
{\sc Gerhard Buchalla and  Oscar Cat\`a} 
\end{center}

\vspace*{0.4cm}

\begin{center}
Ludwig-Maximilians-Universit\"at M\"unchen, Fakult\"at f\"ur Physik,\\
Arnold Sommerfeld Center for Theoretical Physics, 
D--80333 M\"unchen, Germany
\end{center}

\vspace{1.5cm}
\begin{abstract}
\vspace{0.2cm}\noindent
We consider the Standard Model as an effective theory at the weak scale 
$v$ of a generic new strong interaction that dynamically breaks electroweak
symmetry at the energy scale $\Lambda\sim $ (few) TeV. 
Assuming only the minimal field content with the Standard Model
fermions and gauge bosons, but without a light Higgs particle,
we construct the complete Lagrangian through next-to-leading
order, that is, including terms of order $v^2/\Lambda^2$.
The systematics behind this expansion is clarified.
Although similar to chiral perturbation theory, it is not
governed by the dimension of operators alone, but depends
in an essential way on the loop expansion.   
Power-counting formulas are derived that indicate the classes
of operators required at the next-to-leading order.
The complete set of operators at leading and next-to-leading
order is then listed, based on the restrictions implied by
the Standard-Model gauge symmetries. We recover the well-known operators 
discussed in the literature in connection with
the electroweak chiral Lagrangian and in similar contexts,
but we collect a complete and systematic list of all terms
through order $v^2/\Lambda^2$. This includes some operators not
discussed in explicit terms before. We also show that a few of
the previously considered operators can be eliminated via the
equations of motion. As another important result we confirm
the known list of dimension-6 operators in the Standard Model
with an elementary Higgs doublet, essentially as a special
case of our scenario.  
\end{abstract}

\vspace*{2.5cm}

\vfill
\end{titlepage}

\section{Introduction}
\label{sec:intro}

The detection of the $W$ and $Z$ gauge bosons almost thirty years ago 
established that the electroweak interactions are successfully described 
by a $SU(2)_L\otimes U(1)_Y$ gauge group spontaneously broken to $U(1)_Q$. 
However, while the Higgs mechanism is certainly at work, the precise way 
in which the electroweak symmetry is broken remains a mystery. Even whether 
the underlying dynamics is weakly or strongly coupled remains unclear.  

The simplest option for the weakly-coupled scenario is the introduction of 
a fundamental scalar doublet. Besides the required 3 Goldstone bosons, 
one obtains a massive scalar field, the (Standard Model) Higgs boson, 
whose mass has to be taken as a free parameter. Fundamental scalar particles 
however lead to problems of naturalness, and in order to make the theory 
meaningful an additional stabilization mechanism has to be invoked. 
Supersymmetry still stands as the most solid theoretical framework to 
explain the lightness of a weakly-coupled Higgs.  

The alternative is a scenario with dynamical symmetry breaking. 
In this case spontaneous 
symmetry breaking is triggered by a condensate, which is generated by 
new interactions that become strongly coupled at the electroweak 
scale. This scenario is akin to how chiral symmetry is broken in QCD. 
Therefore, one of the distinctive features of strongly-coupled scenarios is 
compositeness and the existence of a large number of bound states, naturally 
starting at the TeV scale. In general, strongly-coupled scenarios allow for 
the presence of a light scalar (with mass around the electroweak scale) if 
it is interpreted as a pseudo-Goldstone boson of a spontaneously broken 
symmetry group~\cite{Kaplan:1983sm,Giudice:2007fh}. The appeal of such models 
is that this (composite) Higgs mass is naturally of the order of the 
electroweak scale~\cite{Georgi:1984ef}.
  
The search for the origin of electroweak symmetry breaking is currently 
underway at the LHC. At the time of writing, ATLAS and CMS have already 
excluded a light Higgs boson for a wide range of masses. 
The non-excluded area is at present in the range 115-130 GeV, with some 
intriguing excess around 125 GeV, not significant enough to be 
conclusive~\cite{ATLAS:2012si,CMS:2012tx}. 
However, even if the existence of a light Higgs 
is confirmed, we will still be unable to discern whether weakly-coupled or 
strongly-coupled scenarios are at work. Additional information at the TeV 
scale will be needed. 
So far, no signals of TeV particles, be it SUSY partners or bound states 
of strongly-coupled theories, have been observed. 

In this article we will study strongly-coupled scenarios. In the past, 
starting with Technicolor models, there has been a huge effort in 
model-building. However, finding a viable UV completion of a strongly-coupled 
scenario has proven a hard task, especially when it comes to match the 
low-energy experimental constraints from LEP. In this paper we will follow 
a model-independent approach. The language we will use is that of Effective 
Field Theories, with the only assumption that the low energy degrees of 
freedom reduce to the presently-established particle content of the 
Standard Model. We will show that such an effective theory is renormalizable 
order by order in a $1/\Lambda^2$ expansion ($\Lambda\sim 4\pi v\sim 3$ TeV), 
and will list the full set of operators up to next-to-leading order (NLO). 
This might be considered the minimal version of an effective theory of 
strongly-coupled electroweak symmetry breaking, in the sense that we are 
including only three Goldstone fields, 
{\it{i.e.}} the longitudinal modes of the $W$ and $Z$ gauge bosons. 

This approach to dynamical electroweak symmetry breaking was first studied 
in~\cite{Appelquist:1980vg}, inspired by the methods developed for chiral 
symmetry breaking in the strong interactions~\cite{Weinberg:1978kz}. Over 
the years there have been different steps towards extending the work of 
\cite{Appelquist:1980vg}, but a systematic classification of operators 
is still lacking. A first analysis of the NLO counterterms for the gauge 
boson and Goldstone sectors was given in~\cite{Longhitano:1980iz}, 
which was followed by a more systematic treatment of the operator basis 
in the CP-conserving~\cite{Longhitano:1980tm} and 
CP-violating~\cite{Appelquist:1993ka} sectors. The interactions of fermionic 
operators with Goldstone modes were first considered 
in~\cite{Appelquist:1984rr,Hauser:1985ey}, where the naive dimension-4 
operators were listed. This list was soon enlarged to include also scalar 
and tensor currents~\cite{Peccei:1989kr,Bagan:1998vu,ManzanoFlecha:2002cx}. 
In the absence of experimental constraints, the coefficients of the different 
operators were initially estimated with specific UV completions, most 
prominently Technicolor-inspired models. With the advent of LEP 1 and 2, 
several phenomenological analyses were devoted to constraining the 
electroweak chiral 
Lagrangian with electroweak precision tests, mostly through oblique 
parameters and triple gauge boson couplings (see, for instance, 
\cite{Dobado:1990zh} and references therein).    

The purpose of this work is twofold: first, we will clarify issues of
power counting and provide a consistent framework for the electroweak 
effective theory. Then we will present a comprehensive classification of 
operators to NLO in the $1/\Lambda^2$ expansion. While in the gauge boson 
sector the list of operators was settled long ago~\cite{Appelquist:1993ka}, 
in the quark sector a systematic treatment is still needed. For instance, 
Ref.~\cite{Bagan:1998vu} considered fermion bilinears but dismissed operators 
proportional to the fermion masses, which is clearly not a suitable 
approximation for top quark physics. Moreover, a full classification 
of 4-fermion operators has not been considered so far in the literature. 
Finally, due care is paid to eliminate possible redundancies and to reduce the 
number of operators to a minimal basis. In particular, we show that some 
relations between operators with fermions and gauge bosons can be obtained 
by direct application of the equations of motion and integration by parts. 
Those relations have already been pointed out 
in \cite{Nyffeler:1999ap,Grojean:2006nn} and can be of importance in 
the study of triple gauge boson couplings \cite{Grojean:2006nn}.

As already mentioned, the framework we introduce in this article assumes 
a minimal particle content for the Standard Model. 
It would be interesting to consider extensions that include a (composite) 
light (pseudo)scalar, if LHC eventually unveils its existence. These 
extensions would share some similarities with the setting presented in 
\cite{Giudice:2007fh}.

The present article is organized as follows: in Section~\ref{sec:lsmlo} 
we review the leading order Lagrangian for the electroweak interactions in 
the absence of a light fundamental Higgs boson. The power counting for the 
effective theory is discussed in Section~\ref{sec:pcount} and then used 
in Section~\ref{sec:lsmnlo} to provide the full list of operators at 
next-to-leading order. As a simple illustration of how such operators can be 
generated from a UV-complete theory, in Section~\ref{sec:smhh} we consider 
as a toy model the Standard Model with a heavy Higgs. In 
Section~\ref{sec:smlh} we contrast our basis of operators with the 
corresponding NLO operators in the presence of a light Higgs boson. 
Related issues of decoupling and renormalizability are discussed 
in Section~\ref{sec:renorm}. Conclusions are given in Section~\ref{sec:concl}. 
The power-counting formulas are derived in Appendix A.
Technical details on the construction of the operator basis at NLO are 
provided in Appendix B, while Appendix C lists the full set of operators in 
the unitary gauge.


\section{SM effective Lagrangian at leading order}
\label{sec:lsmlo}

The field content of the Standard Model (SM) is specified by the 
left-handed doublets $q$, $l$ and right-handed singlets $u$, $d$, $e$
of quarks and leptons, together with the gauge fields $G$, $W$, $B$
of $SU(3)_C$, $SU(2)_L$ and $U(1)_Y$. The fermion fields are
\begin{equation}\label{ql}
q(3,2,1/6),\qquad l(1,2,-1/2),
\qquad u(3,1,2/3), \qquad d(3,1,-1/3),  \qquad e(1,1,-1)
\end{equation}
with their transformation properties under $SU(3)$ and $SU(2)$
and their hypercharge $Y$ indicated in brackets.
Each of the fields carries a generation index $p=1$, $2$, $3$,
which has been suppressed in (\ref{ql}).

The effective theory of the gauge and fermion fields is constructed by
writing down all possible Lorentz-invariant operators, composed of 
these fields, that are singlets under the SM gauge group.
Restriction to operators of dimension less or equal to 4 uniquely
determines the renormalizable, unbroken part of the SM Lagrangian
\begin{eqnarray}\label{lsm4}
{\cal L}_4 &=& -\frac{1}{2} \langle G_{\mu\nu}G^{\mu\nu}\rangle
-\frac{1}{2}\langle W_{\mu\nu}W^{\mu\nu}\rangle 
-\frac{1}{4} B_{\mu\nu}B^{\mu\nu}
\nonumber\\
&& +\bar q i\!\not\!\! Dq +\bar l i\!\not\!\! Dl
 +\bar u i\!\not\!\! Du +\bar d i\!\not\!\! Dd +\bar e i\!\not\!\! De 
\end{eqnarray}
Here and in the following the trace of a matrix $M$ is denoted by
$\langle M\rangle$.
The covariant derivative of a fermion field $\psi_{L,R}$ is defined as
\begin{equation}\label{dcovf}
D_\mu\psi_L =
\partial_\mu \psi_L +i g W_\mu \psi_L +i g' Y_{\psi_L} B_\mu \psi_L , 
\qquad
D_\mu\psi_R =\partial_\mu \psi_R  + i g' Y_{\psi_R} B_\mu \psi_R
\end{equation}
The fields are normalized in the canonical way and the fermionic
terms can always be chosen to be diagonal in the generation index.
We have not written the topological terms $\l G\tilde G\r$ and
$\l W\tilde W\r$. The effective Lagrangian (\ref{lsm4}) describes 
physics at the electroweak scale $v=246\,{\rm GeV}$, assumed to be 
small in comparison with a new physics scale $\Lambda$.

Next, $SU(2)_L\otimes U(1)_Y$ is spontaneously broken
to the electromagnetic $U(1)_Q$. The associated Goldstone bosons
can be described by the nonlinear $SU(2)$ matrix field
\begin{equation}\label{uudef}
U=\exp(2i\Phi/v),\qquad
\Phi=\varphi^a T^a=\frac{1}{\sqrt{2}}\left(
\begin{array}{cc}
\frac{\varphi^0}{\sqrt{2}} & \varphi^+\\
\varphi^- & -\frac{\varphi^0}{\sqrt{2}} 
\end{array}\right)
\end{equation}
with $T^a=T_a$ the generators of $SU(2)$.
We further suppose that no additional light fields exist,
in particular no physical Higgs boson.
Assuming that the Goldstone bosons arise from the spontaneous
breaking of a global $SU(2)_L\otimes SU(2)_R$ symmetry to the diagonal
subgroup $SU(2)_V$, the field $U$ transforms as
\begin{equation}\label{uglgr}
U\rightarrow g_L U g^\dagger_R,\qquad g_{L,R}\in SU(2)_{L,R}
\end{equation}
The transformations $g_L$ and the $U(1)_Y$ subgroup of $g_R$
are gauged, so that the covariant derivative of $U$ is given by
\begin{equation}\label{dcovu}
D_\mu U=\partial_\mu U+i g W_\mu U -i g' B_\mu U T_3
\end{equation}
The dynamics of the Goldstone bosons can be described in a 
model-independent way by constructing the most general Lagrangian,
built from $U$ and the remaining SM fields, that is consistent
with the SM gauge symmetry. In practice this construction is
carried out including all operators that contribute up to a given
order in $v/\Lambda$, where $\Lambda$ is the effective theory cutoff.
This procedure is well-known from chiral perturbation 
theory \cite{Weinberg:1968de}.
The systematics of this expansion has some peculiar features in
the case of the electroweak Standard Model and will be discussed
in detail in the next section. 
Presently we summarize the terms to be included at lowest order
in the Lagrangian, that is, terms of the same order in $v/\Lambda$
as the renormalizable part in (\ref{lsm4}).  

To lowest order the Lagrangian describing the Goldstone bosons
and their interactions with the SM fields is given by
\begin{equation}\label{lulo}
{\cal L}_U = \frac{v^2}{4}\ \l D_\mu U^\dagger D^\mu U\r
-v \left( \bar qY_uU P_+r + 
\bar qY_dU P_-r +\bar lY_eU P_-\eta + {\rm h.c.}\right)
\end{equation}
where the right-handed quark and lepton fields are collected in
$r=(u,d)^T$ and $\eta=(\nu,e)^T$, respectively. As before, we do not
write generation indices explicitly. The Yukawa matrices $Y_{u,d,e}$
are understood to denote the usual matrices in generation space.
We define 
\begin{equation}\label{tpm12def}
P_\pm\equiv \frac{1}{2}\pm T_3\, ,\qquad 
P_{12}\equiv T_1+i T_2\, ,\qquad P_{21}\equiv T_1-i T_2
\end{equation}
where $P_{12}$ and $P_{21}$ will be needed below. 

The first expression in (\ref{lulo}) is the leading term, of ${\cal O}(p^2)$,
in chiral perturbation theory. It contains the kinetic term of the
Goldstone fields and the mass term of the weak gauge bosons. It is
therefore of the same size $\sim v^4$ as the terms in (\ref{lsm4}) and
has to be counted as a leading order contribution.  
In principle, all terms of ${\cal O}(p^2)$ that are invariant
under the SM gauge group $SU(2)_L\otimes U(1)_Y$ have to be included 
in the effective theory at lowest order. 
There are only two such terms, the first operator in (\ref{lulo}) and 
\begin{equation}\label{lub1}
{\cal L}_{\beta_1}=\beta_1 v^2 \langle U^\dagger D_\mu U T_3\rangle^2 
\end{equation}
In contrast to the first term in (\ref{lulo}), the operator in (\ref{lub1}) 
does not respect the global custodial $SU(2)_L\otimes SU(2)_R$ symmetry
\cite{Longhitano:1980iz,Appelquist:1993ka}. It contributes to the 
electroweak $T$-parameter \cite{Peskin:1991sw}, which is strongly
restricted by experimental data. This implies that the coefficient
$\beta_1$ is constrained to be very small ($\beta_1\ll 1$) and suggests
that the custodial symmetry is largely respected by the underlying dynamics 
of electroweak symmetry breaking. If this is the case, the operator
(\ref{lub1}) can in general still be induced through loop corrections.  
Then $\beta_1 v^2\sim v^2/(16\pi^2)\sim v^4/\Lambda^2$ and the contribution
in (\ref{lub1}) amounts to a next-to-leading order correction. 
The related power-counting will be discussed in detail in 
section \ref{sec:pcount}. We have listed this term here for completeness.

The last term in (\ref{lulo}) gives masses to the SM fermions.
Motivated by the case of the top quark, we will count the Yukawa couplings 
$Y_f$ as order one. The fermion mass term then contributes
at the same order as the previous terms. The operators in this term are  
the most general expressions built from fermion bilinears and $U$ fields
without derivatives.
The Lagrangian ${\cal L}_4+{\cal L}_U$ thus defines 
the leading order of the effective theory for the minimal Standard Model 
without a light Higgs boson.

At the quantum level, the previous list of leading order operators has to 
be supplemented with the usual gauge-fixing and Faddeev-Popov terms. 
Expressions for both of these terms in the covariant $R_{\xi}$ gauges are given 
in~\cite{Herrero:1993nc}. In this paper, following~\cite{Appelquist:1980vg}, 
we will work in the Landau gauge. The main advantage of the Landau gauge 
for the electroweak effective theory we are considering is that ghost and 
Goldstone fields decouple. As a result, there is no need to include ghosts 
in the effective operator basis: only Goldstones coupled to gauge and 
fermion fields are needed to renormalize the divergences of the leading-order 
Lagrangian of Eq.~(\ref{lulo}). Additionally, because of this decoupling, 
the ghost term in Landau gauge is exactly the one of the Standard Model. 
Therefore, the presence of the conventional Faddeev-Popov term should be 
implicitly understood in Eq.~(\ref{lulo}).

\section{Power counting}
\label{sec:pcount}

In this section we discuss the systematics of the expansion
that defines the effective theory. This expansion cannot be simply 
organized in terms of the dimension of operators alone. 
This is clear from the Lagrangian ${\cal L}_4+{\cal L}_U$ 
in (\ref{lsm4}) and (\ref{lulo}), which contains operators of
dimension 2, 3 and 4, all contributing at leading order. 
Also, while (\ref{lsm4}) is renormalizable, (\ref{lulo})
contains interactions that are non-renormalizable (in the traditional sense).
Such a theory can still be renormalized (in the modern sense), order
by order in a loop expansion, at the expense of introducing at each order
a finite number of new counterterms, whose structure is dictated by
symmetry. A similar discussion has been given previously in the context
of the chiral quark model in \cite{Manohar:1983md}.

It will be our goal to construct the complete next-to-leading order
Lagrangian supplementing the leading order terms in ${\cal L}_4+{\cal L}_U$.
The next-to-leading order is understood here to include all
operators, allowed by the relevant symmetries, with coefficients
of order $v^2/\Lambda^2$ (or $p^2/\Lambda^2$).
A minimal set of operators is defined by the terms required to absorb
the ultraviolet divergences generated at one loop by the lowest
order Lagrangian ${\cal L}_4+{\cal L}_U$.

\begin{figure}[t]
\begin{center}
\resizebox{4cm}{!}{\includegraphics{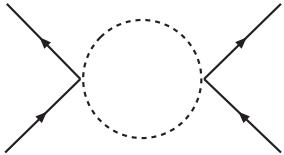}}
\caption{\label{fig:pcexmp}
Goldstone-boson loop contributing to fermion-fermion scattering
induced by the Yukawa interaction in eq. (\ref{lulo}).} 
\end{center}
\end{figure}

As an example, we may consider the terms in the Lagrangian (\ref{lulo}),
which induce at one loop the diagram for $\psi_R\psi_R\to\psi_L\psi_L$
scattering shown in Fig. \ref{fig:pcexmp}.
Each vertex $\bar\psi_L\psi_R \varphi^2$ has a factor of $y/v$,
with $y$ a generic Yukawa coupling. The loop integration produces
a factor of $1/(16\pi^2)=v^2/\Lambda^2$, and the four powers of
loop momentum $p$ in the numerator are compensated by the four powers
of $p$ from the two boson propagators. This leads to the following
schematic expression for the diagram
\begin{equation}
D_1\sim \frac{y^2}{\Lambda^2} p^0\, \bar\psi_L\psi_R\bar\psi_L\psi_R
\end{equation}
This result indicates a 4-fermion amplitude with a logarithmic 
divergence by power counting. The amplitude matches onto a local,
gauge-invariant, dimension-6 operator 
$\bar\psi_L U\psi_R\bar\psi_L U\psi_R$, whose coefficient will
absorb the ultraviolet divergence. We see that this new operator appears
with a natural suppression of order $1/\Lambda^2$. It is therefore part
of the next-to-leading order Lagrangian we want to construct.

The example we have just discussed can be generalized.
Let us in a first step restrict our attention to the Goldstone-fermion
sector, neglecting the gauge fields. This corresponds to taking
the leading order Lagrangian ${\cal L}_4+{\cal L}_U$ in the gaugeless
limit ($g_s$, $g$, $g'\to 0$). In this case the interaction vertices are 
either couplings of an even number $2i$ of Goldstone bosons with
two powers of momentum, contributing a factor of $p^2/v^{2i-2}$,
or couplings of $\bar\psi_L\psi_R$ (and ${\rm h.c.}$) to an arbitrary
number $k$ of bosons, contributing a factor of $y/v^{k-1}$.
Considering a general diagram ${\cal D}$ with these leading-order vertices, 
but with any number of loops $L$ and an arbitrary number of external lines, 
we find the power counting formula 
\begin{equation}\label{pcupsi}
{\cal D}\sim \frac{(yv)^\nu}{v^{F_L+F_R-2}}
  \frac{p^{2L+2-\nu-(F_L+F_R)/2}}{\Lambda^{2L}}
 \bar\psi_L^{F^1_L} \psi_L^{F^2_L} \bar\psi_R^{F^1_R} \psi_R^{F^2_R}
 \left(\frac{\varphi}{v}\right)^B
\end{equation}
where $F_{L,R}=F^1_{L,R}+F^2_{L,R}$ is the number of external fermions
$\psi_{L,R}$, $B$ is the number of external Goldstone bosons $\varphi$ and
$\nu=\sum_k\nu_k$ is the total number of Yukawa vertices, with
$\nu_k$ the number of such vertices that have $k$ boson lines. 
Eq. (\ref{pcupsi}) can be obtained in the standard way by writing
down the powers of $y$, $p$, $v$ and $\Lambda$ for a general diagram
and by eliminating the numbers of internal lines using the familiar
topological identities (see Appendix \ref{sec:apppc}). 
Taking $F_{L,R}=\nu=0$, (\ref{pcupsi}) reduces to
the corresponding expression for pure chiral perturbation theory
\begin{equation}\label{pcchipt}
{\cal D}\sim 
\frac{v^2}{\Lambda^{2L}} p^{2L+2} \left(\frac{\varphi}{v}\right)^B
\end{equation}
This formula encodes the well-known results on the systematics of
chiral perturbation theory: 1) $L$-loop diagrams with only leading-order
vertices contribute at $L$th order in the expansion parameter $p^2/\Lambda^2$
(with $p\sim v$). 2) Since the power of $p$ is $2L+2 > 0$, an 
$L$-loop contribution of this type is divergent and requires an
order $p^{2L+2}$ counterterm. 
According to this pattern only order-$p^2$ terms are required
at the leading order ($L=0$) in the chiral Lagrangian. 
At next-to-leading order ($L=1$) we need in addition
all terms of order $p^4$. These are the $SU(2)_L\otimes U(1)_Y$
gauge invariant operators built from $U$ fields and containing
four covariant derivatives, symbolically of the form $UD^4$.

When fermions are taken into account in addition to the Goldstone
bosons, (\ref{pcupsi}) is the relevant power-counting formula.
To find the required counterterms we proceed as before. 
The order in the expansion in $1/\Lambda^2$ is again determined
by $L$. A non-negative power of $p$ signals a divergent diagram.
At the same time it gives the number of derivatives in the local 
operator that is needed to absorb the divergence. (We are assuming
the use of dimensional regularization throughout the present paper.) 
For the next-to-leading order we are mainly interested in, we have
$L=1$ and the condition for a counterterm reads
\begin{equation}\label{ctcond}
4-\nu-\frac{F_L+F_R}{2}\geq 0
\end{equation}
Since the number of external fermions $F\equiv F_L+F_R$, and also the
number of fermion vertices $\nu$, enters with a negative sign,
the number of fermion fields that can appear in the counterterms
is limited. Eq. (\ref{ctcond}) implies that only $F=2$ and
$F=4$ are allowed. For $F=6$ (and higher) we need at least $\nu=3$ 
Yukawa vertices and (\ref{ctcond}) can no longer be fulfilled.
In these cases, the negative power of $p$ indicates an ultraviolet
convergent diagram, which is part of the finite radiative corrections
predicted at one loop by the lowest order Lagrangian.

The cases that lead to counterterms are then
\begin{equation}\label{fnulist}
(F,\nu,4-\nu-F/2)= (2,1,2),\qquad (2,2,1),\qquad (2,3,0),\qquad (4,2,0)
\end{equation}
The case $(2,3,0)$ corresponds to fermion bilinears with $U$ fields
but without derivatives. These are precisely the Yukawa terms
already present in the leading-order Lagrangian (\ref{lulo}).
The new terms are fermion bilinears with one
or two derivatives, $(2,2,1)$, $(2,1,2)$, and 4-fermion operators with no
derivatives, $(4,2,0)$, all of them with an arbitrary number of $U$ fields.
We will denote these three classes schematically as $\psi^2 UD$, 
$\psi^2 UD^2$ and $\psi^4 U$.

These considerations may be generalized to include interactions with 
gauge fields. In what follows we will omit ghost fields, which would insure
manifest gauge independence, but would not affect the power counting.  
A generic diagram ${\cal D}$ then contains, besides the 
$n_i$ $\varphi^{2i}$-vertices and $\nu_k$ Yukawa interactions 
$\bar\psi_{L(R)}\psi_{R(L)}\varphi^k$, a number $m_l$ of gauge-boson-Goldstone  
vertices $X_\mu \varphi^l$, $r_s$ such vertices of the type
$X^2_\mu  \varphi^s$, $x$ quartic gauge-boson vertices $X^4_\mu$, 
$u$ triple-gauge-boson vertices $X^3_\mu$, and $z_L$ ($z_R$)
fermion-gauge-boson interactions $\bar\psi_{L(R)}\psi_{L(R)}X_\mu$. 
Here $\psi_L$ ($\psi_R$), $\varphi$ and $X_\mu$ denote
left-handed (right-handed) fermions, Goldstone bosons and
gauge fields, respectively.
As discussed in Appendix \ref{sec:apppc}, the power-counting for
a diagram ${\cal D}$ can be summarized by the formula 
\begin{equation}\label{pcupsix}
{\cal D}\sim 
\frac{(yv)^\nu (gv)^{m+2r+2x+u+z}}{v^{F_L+F_R-2}} \frac{p^d}{\Lambda^{2L}}\
\bar\psi_L^{F^1_L} \psi_L^{F^2_L} \bar\psi_R^{F^1_R} \psi_R^{F^2_R}\
\left(\frac{X_{\mu\nu}}{v}\right)^V\  \left(\frac{\varphi}{v}\right)^B
\end{equation}
where the power of $p$ is
\begin{equation}\label{powerd}
d\equiv 2L+2-\frac{F_L+F_R}{2}-V-\nu-m-2r-2x-u-z
\end{equation}
Here $V$ is the number of external gauge-boson lines, $g$ a generic 
gauge coupling, and we have used $\nu\equiv\sum_k\nu_k$, 
$m\equiv\sum_l m_l$, $r\equiv\sum_s r_s$, $z\equiv z_L+z_R$.
Eq. (\ref{pcupsix}) generalizes (\ref{pcupsi}), to which it reduces
when neglecting the gauge fields, that is, taking $V=m=r=x=u=z=0$.
The expression (\ref{powerd}) for $d$ is useful, because
$F_L$, $F_R$ and $V$, as well as the numbers of vertices, all enter 
with a negative sign. This implies a limit to the number of
divergent diagrams at a given order in $L$, 
similar to the special case of (\ref{pcupsi}).
We also note that, as in (\ref{pcupsi}), the number $B$ of external Goldstone 
bosons enters the power counting formula (\ref{pcupsix}) only
through the factor $(\varphi/v)^B$. It is irrelevant in particular for the 
exponent $d$, which counts the powers of momentum. This indicates explicitly
that, at any given order in the effective theory, the counterterms
contain an arbitrary number of Goldstone fields $U=U(\varphi/v)$.

We will next use (\ref{pcupsix}) and (\ref{powerd}) to find the remaining 
classes of NLO counterterms, which contain at least one gauge-field operator.
The cases to be considered are then $L=1$, $F\geq 0$, $V\geq 1$.
\begin{description}
\item $F=0$, $V=1$: In this case we have 
\begin{equation}
d=3-\nu-m-2r-2x-u-z
\end{equation}
Since $V=1$, the diagram must contain at least one gauge-field vertex,
that is, at least one of ($m$, $r$, $x$, $u$, $z$) must be non-zero.
This implies $d\leq 2$. The divergent amplitudes then have $d=2$, $1$ or $0$,
and correspond to counterterms of the form $XUD^2$, $XUD$ and $XU$, 
respectively. No Lorentz scalars can be built from the latter two, 
which leaves us with the operator class $XUD^2$.  
\item $F=0$, $V=2$: The exponent of $p$ now reads $d=2-\nu-m-2r-2x-u-z$. 
With two external gauge fields the diagram has at least two 
single-gauge-field vertices ($m+u+z\geq 2$), or one double-gauge-field vertex 
($r+x\geq 1$). It follows that $d\leq 0$. For $d=0$ the diagram requires a
counterterm, with $F=0$, $V=2$ and $d=0$ derivatives. This determines
its form to be $X^2U$.
\item $F=0$, $V=3$: With $d=1-\nu-m-2r-2x-u-z$ and at least two
vertices required to form a loop diagram, we have $d<0$.
The diagrams of this type are convergent. Operators of class
$X^3$ are therefore not required as counterterms of the nonrenormalizable
effective Lagrangian. Representing operators of dimension 6 they
could a priori still give (finite) contributions of order $1/\Lambda^2$,
that is at NLO. We will return to this and similar cases below.
The case of $F=0$, $V>3$ leads to $d<0$ and to operators beyond
the NLO approximation.
\item $F=2$, $V=1$: Here the power-counting formula gives
$d=2-\nu-m-2r-2x-u-z$. The minimum number of vertices one might
consider is two. This can be realized when a $\psi^2\varphi^2$-vertex
is connected to an $X_\mu\varphi^2$-vertex, contracting the $\varphi$ 
fields to form a one-loop diagram. By inspection, such a diagram is found
to be zero. A nonvanishing diagram thus requires at least three vertices,
implying $d<0$. The diagram converges and the corresponding operators 
of class $\psi^2 UX$ are not needed as counterterms. 
It is clear from these considerations that even larger numbers for 
$F$ and $V$ can only lead to $d < 0$ and to terms beyond the NLO level. 
\end{description}

This completes the enumeration of NLO counterterms.
To summarize, power counting allows us to identify the classes of
operators we need to include in the next-to-leading order Lagrangian.
Denoting by $U$ the presence of any number of Goldstone fields
$U$ (or $U^\dagger$), and by $D^n$, $\psi^F$, $X^k$ the numbers
$n$, $F$, $k$, respectively, of derivatives, fermion fields and gauge-boson
field-strength tensors, these classes are schematically given by
\begin{equation}\label{nloclass}                                         
UD^4,\qquad XUD^2,\qquad X^2 U,\qquad \psi^2 UD,\qquad \psi^2 UD^2,\qquad   
\psi^4 U                                                              
\end{equation}
We remark that all the possible classes of 
dimension-4 operators built from $U$, $D$, $X$ and $\psi$
appear as NLO counterterms ($UD^4$, $XUD^2$, $X^2U$, $\psi^2 UD$), 
but there are also operators of dimension 5 and 6 at the same 
level ($\psi^2 UD^2$, $\psi^4 U$). 

All operators that are not required as counterterms at NLO are
naturally of higher order in $1/\Lambda^2$.  This is clear for
operators of canonical dimension larger than 6. These have a
coefficient $\sim 1/\Lambda^n$, $n\geq 3$ on purely dimensional grounds,
which is possibly further suppressed by loop factors 
$1/(16\pi^2)=v^2/\Lambda^2$. A special consideration is needed for
operators of dimension 5 and 6, since those may appear either at NLO
($\sim 1/\Lambda^2$), or only at NNLO ($\sim 1/\Lambda^4$). The former case 
is illustrated by operators of the type $(\bar\psi_LU\psi_R)^2$.
Examples for the latter case are operators with Goldstone fields
and six derivatives (class $UD^6$). 
More systematically, the various classes of such operators of 
dimension 5 and 6, which would only appear as counterterms at NNLO,
are given by:
$UD^6$, $XUD^4$, $X^2UD^2$, $X^3U$, $\psi^2 UD^3$, $\psi^2 UXD$,
and $\psi^2 UX$.  
We will not provide a complete classification of those operators, 
but we would like to comment on a few relevant aspects.    

To some extent the actual size of the coefficient of a given operator
depends on the UV completion of the theory. It is, however, natural to 
expect that a NNLO counterterm will have a coefficient no larger than
${\cal O}(1/\Lambda^4)$. As an illustration we may consider the operator
$G_{\mu\nu}G^{\mu\nu}\langle D_\lambda U^\dagger D^\lambda U\rangle$, which
belongs to the class $X^2UD^2$ of NNLO counterterms. In a model with a 
heavy Higgs boson, this operator would be generated by Higgs exchange
between the Higgs-Goldstone couplings and the Higgs-gluon-gluon vertex.
The latter would contain a loop factor $v^2/\Lambda^2$, while the Higgs
propagator would contribute a factor of $1/m^2_h\sim 1/\Lambda^2$,
leading to an ${\cal O}(1/\Lambda^4)$ coefficient. A similar consideration 
applies to the operators $\psi^2 UX$ and $X^3U$, which are typically 
loop induced \cite{Arzt:1994gp}. The NNLO classes $\psi^2 UX$ and $X^3U$
are special in that they directly correspond to dimension-6 operators
appearing at NLO in the conventional Standard Model with a linear Higgs
doublet. Since we want to compare this scenario with our framework
later on, we will also list below the few operators of classes  
$\psi^2 UX$ and $X^3$ (but not $X^3U$).

As we have seen above, 4-fermion
operators of the form $\bar\psi_L U\psi_R\, \bar\psi_L U\psi_R$
arise as counterterms at next-to-leading order in the effective
Lagrangian and thus have to be included in our operator basis.
We are thus led to consider the entire class of 4-fermion operators,
which may include an arbitrary number of $U$ fields. However,
not all of these operators are required as counterterms of the
Lagrangian ${\cal L}_4+{\cal L}_U$ at next-to-leading order.
An example is the case of
$\bar\psi_L\gamma^\mu \psi_L\, \bar\psi_L \gamma_\mu\psi_L$,
whose one-loop coefficient is finite by power counting
($F_R=0$, $F_L=\nu=4$ in (\ref{ctcond})). It is then natural
to expect this term only at NNLO $\sim 1/\Lambda^4$.
However, depending on the details of the physics at the $TeV$ scale,
such an operator could possibly be generated through tree-level
exchange of a heavy boson. In this case, the coefficient could be
of order $1/\Lambda^2$ and the operator would enter already at NLO,
giving a finite contribution at this level.
We will not distinguish explicitly between NLO and NNLO counterterms in the 
sector $\psi^4 U$ and include the complete list of 4-fermion operators
in our basis, except those that violate baryon ($B$) or lepton number ($L$).
We will briefly return to such operators in section~\ref{subsec:blvop}.

We finally remark that, independent of the nonrenormalizable interactions
organized into a $1/\Lambda^2$ expansion, the usual corrections from the 
renormalizable part of the Standard Model can be computed, in principle
to arbitrary order. This applies e.g. to corrections from perturbative
QCD. 

In the following section we will provide the complete list of NLO operators
for each of the classes in (\ref{nloclass}).
As discussed above, we will also consider the operators
$X^3$ and $\psi^2 UX$.

\section{SM effective Lagrangian at next-to-leading order}
\label{sec:lsmnlo}

The next-to-leading-order effective Lagrangian of the Standard Model 
with dynamically broken electroweak symmetry and minimal particle content 
can be written as
\begin{equation}\label{leffnlo}
{\cal L}={\cal L}_4+{\cal L}_U+{\cal L}_{\beta_1} +
\sum_i c_i\frac{v^{6-d_i}}{\Lambda^2}\, {\cal O}_i
\end{equation}
Here ${\cal L}_4+{\cal L}_U$ is the leading order Lagrangian
and ${\cal L}_{\beta_1}$ the custodial-symmetry breaking, 
dimension-2 operator of section \ref{sec:lsmlo}. As discussed there,
${\cal L}_{\beta_1}$ can be treated as an NLO correction.
Apart from this term, the full set of NLO operators is
denoted by the ${\cal O}_i$ in (\ref{leffnlo}). They come with a
suppression by two powers of the cutoff $\Lambda\approx 4\pi v$
and have dimensionless coefficients $c_i$, which are naturally of order 
unity. $d_i$ is the canonical dimension of the operator ${\cal O}_i$.
We will in general assume conservation of baryon and lepton number 
($B$ and $L$) in (\ref{leffnlo}) and return to a brief consideration 
of $B$ and $L$ violating effects in section \ref{subsec:blvop}.

In the following we list all NLO operators ${\cal O}_i$, grouped
according to the subsets introduced at the end of section \ref{sec:pcount}.
Most of the operators are hermitean. For the others, independent hermitean 
conjugate versions are always understood and have not been listed separately.
We will comment on some important points here but will otherwise 
relegate details on the derivation to Appendix \ref{sec:appa}. 

\subsection{\boldmath $UD^4$ terms}

The operators of this class correspond to the ${\cal O}(p^4)$ 
chiral-Lagrangian terms already listed by Longhitano \cite{Longhitano:1980tm}.
They read

\begin{eqnarray}  
{\cal O}_{D1} &=& \l D_\mu U^\dagger D^\mu U\r^2 \nonumber\\
{\cal O}_{D2} &=& \l D_\mu U^\dagger D_\nu U\r
  \ \l D^\mu U^\dagger D^\nu U\r\nonumber\\
{\cal O}_{D3} &=& \left(\l U^\dagger D_\mu UT_3\r\
   \l U^\dagger D^\mu UT_3 \r\right)^2\nonumber\\ 
{\cal O}_{D4} &=& \l U^\dagger D_\mu UT_3\r\ \l U^\dagger D^\mu UT_3\r 
  \ \l D_\nu U^\dagger D^\nu U\r\nonumber\\
{\cal O}_{D5} &=& \l U^\dagger D_\mu UT_3\r\  \l U^\dagger D_\nu UT_3\r 
  \  \l D^\mu U^\dagger D^\nu U\r
\end{eqnarray}

\subsection{\boldmath $XUD^2$ and $X^2U$ terms}

The CP-even operators are \cite{Longhitano:1980tm,Appelquist:1993ka}
\begin{eqnarray}\label{xudev}  
{\cal O}_{XU1} &=& g'g\ B_{\mu\nu}\ \l U^\dagger W^{\mu\nu} U T_3\r \nonumber\\
{\cal O}_{XU2} &=& g^2\ \l U^\dagger W_{\mu\nu} U T_3\r\
   \l U^\dagger W^{\mu\nu} U T_3\r \nonumber\\
{\cal O}_{XU3} &=& g\ \varepsilon^{\mu\nu\lambda\rho}\
   \l U^\dagger W_{\mu\nu} D_\lambda U\r\ 
   \l U^\dagger D_\rho U T_3\r
\end{eqnarray}

The list of $X^2U$ and $XUD^2$ operators also includes
three independent CP-odd terms \cite{Appelquist:1993ka}
\begin{eqnarray}\label{xudod}                                              
{\cal O}_{XU4} &=& g' g\ \varepsilon^{\mu\nu\lambda\rho}\
  B_{\mu\nu}\ \l U^\dagger W_{\lambda\rho} U T_3\r \nonumber\\          
{\cal O}_{XU5} &=& g^2\ \varepsilon^{\mu\nu\lambda\rho}\ 
\l U^\dagger W_{\mu\nu} U T_3\r\ \l U^\dagger W_{\lambda\rho} U T_3\r \nonumber\\  
{\cal O}_{XU6} &=& g\ \l U^\dagger W_{\mu\nu} D^\mu U\r\ 
\l U^\dagger D^\nu U T_3\r 
\end{eqnarray}

We note that three more CP-even operators that had been listed in 
\cite{Appelquist:1993ka} can be expressed in terms of other operators in the 
complete basis and are thus redundant \cite{Nyffeler:1999ap,Grojean:2006nn}. 
This follows from applying the equations of motion and discarding total 
derivatives. For example, the first of these operators is
\begin{eqnarray}\label{xudred1}  
&& B_{\mu\nu}\ \l D^\mu U^\dagger D^\nu U T_3\r \equiv \\
&&\partial^\mu\left[B_{\mu\nu}\l U^\dagger D^\nu UT_3\r \right]
-\partial^\mu B_{\mu\nu} \l U^\dagger D^\nu UT_3\r 
-\frac{1}{2} B_{\mu\nu}  \l U^\dagger(D^\mu D^\nu-D^\nu D^\mu)UT_3\r = \nonumber\\
&& \frac{i g'}{2} v^2 \l U^\dagger D^\nu UT_3\r^2 -
g' \sum_\psi Y_\psi \bar\psi\gamma_\nu\psi\, \l U^\dagger D^\nu UT_3\r
-\frac{ig}{2} B_{\mu\nu}\, \l U^\dagger W^{\mu\nu} UT_3\r
+\frac{i g'}{4} B_{\mu\nu} B^{\mu\nu} \nonumber
\end{eqnarray}
where the second equality holds up to the total derivative.
Similar considerations apply to the operators
\begin{equation}\label{xudred23}
\l U^\dagger W_{\mu\nu} U T_3\r\ \l D^\mu U^\dagger D^\nu U T_3\r\, ,\qquad 
\l U^\dagger W_{\mu\nu} U D^\mu U^\dagger D^\nu U\r 
\end{equation}
as will be further discussed in Appendix \ref{sec:appa}.

\subsection{\boldmath $\psi^2UD$ terms}

For the quark and lepton sectors one finds in this class:

\begin{equation}\label{psi2d}
\begin{array}{cc}
{\cal O}_{\psi V1}=\bar q\gamma^\mu q\ \l U^\dagger iD_\mu UT_3\r ,\qquad &
{\cal O}_{\psi V4}=\bar u\gamma^\mu u\ \l U^\dagger iD_\mu UT_3\r 
\vspace*{0.2cm}\\
{\cal O}_{\psi V2}=\bar q\gamma^\mu UT_3U^\dagger q\ 
\l U^\dagger iD_\mu UT_3\r , \qquad &
{\cal O}_{\psi V5}=\bar d\gamma^\mu d\ \l U^\dagger iD_\mu UT_3\r 
\vspace*{0.2cm}\\
{\cal O}_{\psi V3}=\bar q\gamma^\mu U P_{12} U^\dagger q\ 
\l U^\dagger iD_\mu U P_{21}\r \quad ({\rm h.c.}), \qquad &
{\cal O}_{\psi V6}=\bar u\gamma^\mu d\ 
\l U^\dagger iD_\mu UP_{21}\r \quad ({\rm h.c.}) \\
& \\
{\cal O}_{\psi V7}=\bar l\gamma^\mu l\ \l U^\dagger iD_\mu UT_3\r ,\qquad &
{\cal O}_{\psi V10}=\bar e\gamma^\mu e\ \l U^\dagger iD_\mu UT_3\r 
\vspace*{0.2cm}\\
{\cal O}_{\psi V8}=\bar l\gamma^\mu UT_3U^\dagger l\ 
\l U^\dagger iD_\mu UT_3\r \qquad & 
\vspace*{0.2cm}\\
{\cal O}_{\psi V9}=\bar l\gamma^\mu U P_{12} U^\dagger l\ 
\l U^\dagger iD_\mu U P_{21}\r \quad ({\rm h.c.})  \qquad & 
\end{array}
\vspace*{0.2cm}
\end{equation}

These operators are equivalent to those given as
${\cal L}^1_4,\ldots, {\cal L}^6_4$ in eq. (20) of \cite{Appelquist:1984rr}.
The term ${\cal L}^7_4$ listed there can be eliminated using the
equations of motion, as already noted in \cite{Bagan:1998vu}.

In general, ${\cal O}_{\psi V3}$ and its hermitean conjugate are
two independent operators. If the coefficients are real, only the combination 
${\cal O}_{\psi V3} + {\cal O}^\dagger_{\psi V3}$ can appear in the Lagrangian.
This combination can be eliminated in favour of the simpler operator
\begin{equation}\label{qqpv23}
\bar q\gamma^\mu iD_\mu U U^\dagger q \equiv 
{\cal O}_{\psi V3} + {\cal O}^\dagger_{\psi V3} + 2 {\cal O}_{\psi V2}
\end{equation}
Similar considerations apply to ${\cal O}_{\psi V9}$, where
\begin{equation}\label{llpv89}                                             
\bar l\gamma^\mu iD_\mu U U^\dagger l \equiv                                
{\cal O}_{\psi V9} + {\cal O}^\dagger_{\psi V9} + 2 {\cal O}_{\psi V8}     
\end{equation}

\subsection{\boldmath $\psi^2UD^2$ terms}

The operators with fermionic scalar currents are:

\begin{eqnarray}
{\cal O}_{\psi S1,2} &=& \bar q UP_\pm r\ \l D_\mu U^\dagger D^\mu U \r 
\nonumber\\
{\cal O}_{\psi S3,4} &=& \bar q UP_\pm r\ \l U^\dagger D_\mu U T_3\r^2
\nonumber\\
{\cal O}_{\psi S5} &=& \bar q UP_{12} r\ \l U^\dagger D_\mu U P_{21}\r \ 
\l U^\dagger D^\mu U T_3\r \nonumber\\
{\cal O}_{\psi S6} &=& \bar q UP_{21} r\ \l U^\dagger D_\mu U P_{12}\r \ 
\l U^\dagger D^\mu U T_3\r \nonumber\\
{\cal O}_{\psi S7} &=& \bar l UP_- \eta\ \l D_\mu U^\dagger D^\mu U \r 
\nonumber\\
{\cal O}_{\psi S8} &=& \bar l UP_- \eta\ \l U^\dagger D_\mu U T_3\r^2
\nonumber\\
{\cal O}_{\psi S9} &=& \bar l UP_{12} \eta\ \l U^\dagger D_\mu U P_{21}\r \ 
\l U^\dagger D^\mu U T_3\r 
\end{eqnarray}

The operators with fermionic tensor currents are:

\begin{eqnarray}                                                               
{\cal O}_{\psi T1} &=& \bar q\sigma^{\mu\nu} UP_{12} r\ 
\l U^\dagger D_\mu U P_{21}\r \  \l U^\dagger D_\nu U T_3\r \nonumber\\     
{\cal O}_{\psi T2} &=& \bar q\sigma^{\mu\nu} UP_{21} r\ 
\l U^\dagger D_\mu U P_{12}\r \  \l U^\dagger D_\nu U T_3\r \nonumber\\    
{\cal O}_{\psi T3,4} &=& \bar q\sigma^{\mu\nu} UP_\pm r\ 
\l U^\dagger D_\mu U P_{12}\r \  \l U^\dagger D_\nu U P_{21}\r \nonumber\\  
{\cal O}_{\psi T5} &=& \bar l\sigma^{\mu\nu} UP_{12} \eta\ 
\l U^\dagger D_\mu U P_{21}\r \ \l U^\dagger D_\nu U T_3\r \nonumber\\      
{\cal O}_{\psi T6} &=& \bar l\sigma^{\mu\nu} UP_- \eta\ 
\l U^\dagger D_\mu U P_{12}\r \  \l U^\dagger D_\nu U P_{21}\r            
\end{eqnarray}

\subsection{\boldmath $\psi^4U$ and $\psi^4$ terms}

The dimension-6 operators with four fermion fields and an arbitrary
number of Goldstone fields $U$ can be classified as follows. First, there 
are the 4-fermion operators without $U$ fields (class $\psi^4$), 
which have already been listed in \cite{Grzadkowski:2010es}.
In this case the 4 fermions necessarily have vanishing total hypercharge,
$Y(\psi^4)=0$. A new class arises when these operators are dressed with 
zero-hypercharge combinations of $U$ fields that are contracted in all 
possible $SU(2)$-invariant ways with the fermion doublets.
Beyond this possibility we have to consider operators with 4 fermions
that have a combined hypercharge of $Y(\psi^4)=\pm 1/2$, $\pm 1$, $\pm 3/2$
and $\pm 2$, because these values can be compensated by the
inclusion of $U$ fields. Since $U$ carries $Y=\pm 1/2$ and we can have
at most 4 factors of $U$ to form $SU(2)$ invariants with the maximum
number of 4 $SU(2)$-doublet fermions, operators with $|Y(\psi^4)|> 2$
are not allowed. Listing all 4-fermion combinations, it turns out
that the classes $Y(\psi^4)=\pm 1/2$, $\pm 3/2$ and $\pm 2$ contain only
baryon or lepton number violating operators. This leaves us
with $\psi^4U$ operators that either have $Y(\psi^4)=0$ or $\pm 1$.  
In the following all fermion fields carry again an independent generation 
index, which is suppressed in our notation. 
The generators of $SU(2)_L$ ($SU(3)_C$) are denoted by $T^a$ ($T^A$). 
Hermitian conjugate versions of the operators listed below are understood.

The $\psi^4$ operators \cite{Grzadkowski:2010es}
can be divided into four groups:  
$\bar LL\bar LL$, $\bar RR\bar RR$, $\bar LL\bar RR$,
and $\bar LR\bar LR$, according to the handedness of the fermions. 

The $\bar LL\bar LL$ operators are
\begin{equation}\label{llll1}
{\cal O}_{LL1}=\bar q\gamma^\mu q\, \bar q\gamma_\mu q\, ,\qquad
{\cal O}_{LL2}=\bar q\gamma^\mu T^a q\, \bar q\gamma_\mu T^a q
\nonumber
\end{equation}
\begin{equation}\label{llll2}                                              
{\cal O}_{LL3}=\bar q\gamma^\mu q\, \bar l\gamma_\mu l\, ,\qquad          
{\cal O}_{LL4}=\bar q\gamma^\mu T^a q\, \bar l\gamma_\mu T^a l            
\nonumber
\end{equation}
\begin{equation}\label{llll3}                                               
{\cal O}_{LL5}=\bar l\gamma^\mu l\, \bar l\gamma_\mu l                         
\end{equation}

The $\bar RR\bar RR$ operators are
\begin{equation}\label{rrrr1}                                               
{\cal O}_{RR1}=\bar u\gamma^\mu u\, \bar u\gamma_\mu u\, ,\qquad         
{\cal O}_{RR2}=\bar d\gamma^\mu d\, \bar d\gamma_\mu d                    
\nonumber
\end{equation}
\begin{equation}\label{rrrr2}                                                
{\cal O}_{RR3}=\bar u\gamma^\mu u\, \bar d\gamma_\mu d\, ,\qquad            
{\cal O}_{RR4}=\bar u\gamma^\mu T^A u\, \bar d\gamma_\mu T^A d               
\nonumber
\end{equation}
\begin{equation}\label{rrrr3}                                               
{\cal O}_{RR5}=\bar u\gamma^\mu u\, \bar e\gamma_\mu e\, ,\qquad           
{\cal O}_{RR6}=\bar d\gamma^\mu d\, \bar e\gamma_\mu e                    
\nonumber
\end{equation}
\begin{equation}\label{rrrr4}                                               
{\cal O}_{RR7}=\bar e\gamma^\mu e\, \bar e\gamma_\mu e                    
\end{equation}

The $\bar LL\bar RR$ operators are
\begin{equation}\label{llrr1}                                               
{\cal O}_{LR1}=\bar q\gamma^\mu q\, \bar u\gamma_\mu u\, ,\qquad           
{\cal O}_{LR2}=\bar q\gamma^\mu T^A q\, \bar u\gamma_\mu T^A u             
\nonumber
\end{equation}
\begin{equation}\label{llrr2}                                               
{\cal O}_{LR3}=\bar q\gamma^\mu q\, \bar d\gamma_\mu d\, ,\qquad      
{\cal O}_{LR4}=\bar q\gamma^\mu T^A q\, \bar d\gamma_\mu T^A d            
\nonumber
\end{equation}
\begin{equation}\label{llrr3}                                               
{\cal O}_{LR5}=\bar u\gamma^\mu u\, \bar l\gamma_\mu l\, ,\qquad         
{\cal O}_{LR6}=\bar d\gamma^\mu d\, \bar l\gamma_\mu l                  
\nonumber
\end{equation}
\begin{equation}\label{llrr4}                                               
{\cal O}_{LR7}=\bar q\gamma^\mu q\, \bar e\gamma_\mu e\, ,\qquad           
{\cal O}_{LR8}=\bar l\gamma^\mu l\, \bar e\gamma_\mu e
\nonumber
\end{equation}
\begin{equation}\label{llrr5}                                              
{\cal O}_{LR9}=\bar q\gamma^\mu l\, \bar e\gamma_\mu d               
\end{equation}

The $\bar LR\bar LR$ operators are
\begin{equation}\label{lrrl1}                                             
{\cal O}_{ST1}=\varepsilon_{ij}\, \bar q^i u\, \bar q^j d\, ,\qquad
{\cal O}_{ST2}=\varepsilon_{ij}\, \bar q^i T^A u\, \bar q^j T^A d
\nonumber
\end{equation}
\begin{equation}\label{lrrl2}                                             
{\cal O}_{ST3}=\varepsilon_{ij}\, \bar q^i u\, \bar l^j e\, ,\qquad       
{\cal O}_{ST4}=\varepsilon_{ij}\, \bar q^i \sigma^{\mu\nu} u\, 
\bar l^j \sigma_{\mu\nu}  e 
\end{equation}

The $\bar LL\bar LL$ operators with $U$ fields are
($\alpha$, $\beta$ denote colour indices)
\begin{equation}\label{llllu1}
{\cal O}_{LL6}=
\bar q\gamma^\mu UT_3U^\dagger q\, \bar q\gamma_\mu UT_3U^\dagger q\, ,\qquad
{\cal O}_{LL7}=\bar q\gamma^\mu UT_3U^\dagger q\, \bar q\gamma_\mu q
\nonumber
\end{equation}
\begin{equation}\label{llllu2}    
{\cal O}_{LL8}=\bar q_\alpha\gamma^\mu UT_3U^\dagger q_\beta\, 
\bar q_\beta\gamma_\mu UT_3 U^\dagger q_\alpha\, ,\qquad
{\cal O}_{LL9}=\bar q_\alpha\gamma^\mu  UT_3U^\dagger q_\beta\, 
\bar q_\beta\gamma_\mu  q_\alpha
\nonumber
\end{equation}
\begin{equation}\label{llllu3}
{\cal O}_{LL10}=                                              
\bar q\gamma^\mu UT_3U^\dagger q\, \bar l\gamma_\mu UT_3U^\dagger l
\nonumber
\end{equation}
\begin{equation}\label{llllu3b} 
{\cal O}_{LL11}=\bar q\gamma^\mu UT_3U^\dagger q\, \bar l\gamma_\mu l\, ,\qquad
{\cal O}_{LL12}=\bar q\gamma^\mu q\, \bar l\gamma_\mu UT_3U^\dagger l
\nonumber
\end{equation}
\begin{equation}\label{llllu4}    
{\cal O}_{LL13}=\bar q\gamma^\mu  UT_3U^\dagger l\, 
\bar l\gamma_\mu  UT_3U^\dagger q\, ,\qquad  
{\cal O}_{LL14}=\bar q\gamma^\mu  UT_3U^\dagger l\, \bar l\gamma_\mu  q
\nonumber
\end{equation}
\begin{equation}\label{llllu5}     
{\cal O}_{LL15}=                                          
\bar l\gamma^\mu UT_3U^\dagger l\, \bar l\gamma_\mu UT_3U^\dagger l\, ,\qquad 
{\cal O}_{LL16}=\bar l\gamma^\mu UT_3U^\dagger l\, \bar l\gamma_\mu l     
\end{equation}

The $\bar LL\bar RR$ operators with $U$ fields are
\begin{equation}\label{llrru1}                                               
{\cal O}_{LR10}=
\bar q\gamma^\mu UT_3U^\dagger q\, \bar u\gamma_\mu u\, ,\qquad              
{\cal O}_{LR11}=\bar q\gamma^\mu T^A UT_3U^\dagger q\, \bar u\gamma_\mu T^A u
\nonumber
\end{equation}
\begin{equation}\label{llrru2}                                               
{\cal O}_{LR12}=
\bar q\gamma^\mu UT_3U^\dagger q\, \bar d\gamma_\mu d\, ,\qquad               
{\cal O}_{LR13}=\bar q\gamma^\mu T^A UT_3U^\dagger q\, \bar d\gamma_\mu T^A d
\nonumber
\end{equation}
\begin{equation}\label{llrru3}                                               
{\cal O}_{LR14}=
\bar u\gamma^\mu u\, \bar l\gamma_\mu UT_3U^\dagger l\, ,\qquad               
{\cal O}_{LR15}=\bar d\gamma^\mu d\, \bar l\gamma_\mu UT_3U^\dagger l     
\nonumber
\end{equation}
\begin{equation}\label{llrru4}                                               
{\cal O}_{LR16}=
\bar q\gamma^\mu UT_3U^\dagger q\, \bar e\gamma_\mu e\, ,\qquad               
{\cal O}_{LR17}=\bar l\gamma^\mu UT_3U^\dagger l\, \bar e\gamma_\mu e
\nonumber
\end{equation}
\begin{equation}\label{llrru5}                                              
{\cal O}_{LR18}=\bar q\gamma^\mu UT_3U^\dagger l\, \bar e\gamma_\mu d     
\end{equation}

The $\bar LR\bar LR$ operators with $U$ fields are
\begin{equation}\label{lrrlu1}                                             
{\cal O}_{ST5}=\bar q UP_+ r\, \bar q UP_- r\, ,\qquad
{\cal O}_{ST6}=\bar q UP_{21} r\, \bar q UP_{12} r
\nonumber
\end{equation}
\begin{equation}\label{lrrlu2}                                             
{\cal O}_{ST7}=\bar q UP_+ T^A r\, \bar q UP_- T^A r\, ,\qquad
{\cal O}_{ST8}=\bar q UP_{21} T^A r\, \bar q UP_{12} T^A r
\nonumber
\end{equation}
\begin{equation}\label{lrrlu3}                                             
{\cal O}_{ST9}=\bar q UP_+ r\, \bar l UP_- \eta\, ,\qquad
{\cal O}_{ST10}=\bar q UP_{21} r\, \bar l UP_{12} \eta
\nonumber
\end{equation}
\begin{equation}\label{lrrlu4}                                             
{\cal O}_{ST11}=
\bar q\sigma^{\mu\nu} UP_+ r\, \bar l\sigma_{\mu\nu} UP_- \eta\, ,\qquad
{\cal O}_{ST12}=
\bar q\sigma^{\mu\nu} UP_{21} r\, \bar l\sigma_{\mu\nu} UP_{12} \eta
\end{equation}

Finally, the $\psi^4 U$ operators with $Y(\psi^4)=\pm 1$ are
\begin{equation}\label{psi4y11}                                             
{\cal O}_{FY1}=\bar q UP_+ r\, \bar q UP_+ r\, ,\qquad
{\cal O}_{FY2}=\bar q UP_+ T^A r\, \bar q UP_+ T^A r
\nonumber
\end{equation}
\begin{equation}\label{psi4y12}                                             
{\cal O}_{FY3}=\bar q UP_- r\, \bar q UP_- r\, ,\qquad
{\cal O}_{FY4}=\bar q UP_- T^A r\, \bar q UP_- T^A r
\nonumber
\end{equation}
\begin{equation}\label{psi4y13}                                             
{\cal O}_{FY5}=\bar q UP_- r\, \bar r P_+ U^\dagger q\, ,\qquad
{\cal O}_{FY6}=\bar q UP_- T^A r\, \bar r P_+ U^\dagger T^A q
\nonumber
\end{equation}
\begin{equation}\label{psi4y14}                                             
{\cal O}_{FY7}=\bar q UP_- r\, \bar l UP_- \eta\, ,\qquad
{\cal O}_{FY8}=\bar q\sigma^{\mu\nu} UP_- r\, \bar l\sigma_{\mu\nu} UP_- \eta
\nonumber
\end{equation}
\begin{equation}\label{psi4y15}                                             
{\cal O}_{FY9}=\bar l UP_- \eta\, \bar r P_+ U^\dagger q
\nonumber
\end{equation}
\begin{equation}\label{psi4y16}                                             
{\cal O}_{FY10}=\bar l UP_- \eta\, \bar l U P_- \eta
\nonumber
\end{equation}
\begin{equation}\label{psi4y17}                                             
{\cal O}_{FY11}=\bar l UP_- r\, \bar r P_+ U^\dagger l
\end{equation}

\subsection{\boldmath $X^3$ and $\psi^2 UX$ terms}

The operators $X^3$, built from 3 factors
of field-strength tensors, are not required as counterterms at 
next-to-leading order. Being of dimension 6, they are suppressed by 
two powers of the heavy mass scale $\Lambda$. A loop suppression
will bring the coefficients further down to the NNLO level
$\sim 1/\Lambda^4$. Only if they could be induced at tree level
these operators would give (finite) contributions at NLO. 
We will include them for completeness, as discussed in
\ref{sec:pcount}. In class $X^3$, there are only four operators 
\cite{Grzadkowski:2010es,Buchmuller:1985jz}
\begin{equation}\label{x3g}
{\cal O}_{X1}=f^{ABC} G^{A\nu}_\mu G^{B\rho}_\nu G^{C\mu}_\rho\, ,\qquad
{\cal O}_{X2}=f^{ABC} \tilde G^{A\nu}_\mu G^{B\rho}_\nu G^{C\mu}_\rho
\end{equation}
\begin{equation}\label{x3w}                                               
{\cal O}_{X3}=\varepsilon^{abc} W^{a\nu}_\mu W^{b\rho}_\nu W^{c\mu}_\rho\, ,\qquad    
{\cal O}_{X4}=\varepsilon^{abc} \tilde W^{a\nu}_\mu W^{b\rho}_\nu W^{c\mu}_\rho
\end{equation}
where $f^{ABC}$ and $\varepsilon^{abc}$ are the structure constants
of colour $SU(3)$ and weak $SU(2)$, respectively.
The dual field strength is defined by
\begin{equation}\label{xdual}
\tilde X_{\mu\nu}=\frac{1}{2}\varepsilon_{\mu\nu\rho\sigma} X^{\rho\sigma}\, ,
\qquad \varepsilon^{0123}=-1
\end{equation}

Similar comments apply to the dimension-5 operators of
class $\psi^2 UX$. Again there are only a few structures, which 
can be written down as follows:
\begin{equation}
{\cal O}_{\psi X1,2}=g_s\bar q\sigma^{\mu\nu}G_{\mu\nu}UP_\pm r \nonumber
\end{equation}
\begin{equation}
{\cal O}_{\psi X3,4}=g\bar q\sigma^{\mu\nu}W_{\mu\nu}UP_\pm r,\qquad 
{\cal O}_{\psi X5}=g\bar l\sigma^{\mu\nu}W_{\mu\nu}UP_-\eta \nonumber
\end{equation}
\begin{equation}\label{psi2x}
{\cal O}_{\psi X6,7}=g' \bar q\sigma^{\mu\nu}B_{\mu\nu}UP_\pm r,\qquad 
{\cal O}_{\psi X8}=g' \bar l\sigma^{\mu\nu}B_{\mu\nu}UP_-\eta 
\end{equation}

\subsection{Baryon and lepton number violating operators}
\label{subsec:blvop}

The suppression of $B$-violating operators has to be very strong
in view of the stringent lower limit on the proton lifetime, pointing
to the energy scale of grand unified theories (GUT), which could
naturally generate such operators. Such a large scale is unrelated
to the scale of electroweak symmetry breaking. We will therefore not include 
those terms explicitly in our basis. The five $B$ (and $L$) violating
operators built from four Standard-Model fermions are well 
known \cite{Weinberg:1979sa} 
and can also be found in \cite{Grzadkowski:2010es}.

Similarly, $B$-conserving but $L$-violating operators can be considered.
The fundamental scale of lepton-number violation is not well determined
and might in principle range from a few TeV to values that are many
orders of magnitude larger. Since $L$ violation is relevant for a general 
description of neutrino masses, we want to include the dominant
effects of this type in the lowest-order Lagrangian. In our framework
the leading $L$-violating operator has dimension 3 and can be
written as
\begin{equation}\label{lvdim3}
Q_{LV}=l^T C\, U^* P_+ U^\dagger l
\end{equation}
where $C=i\gamma^2\gamma^0$ is the charge-conjugation matrix.
A similar operator has already been considered in \cite{Hirn:2005fr}.
$Q_{LV}$ is unique up to the various possible flavour assignments of $l$.
This is in full analogy to the well-known $L$-violating dimension-5 
operator in the usual Standard Model
\begin{equation}\label{lvdim5}                                              
(\tilde\phi^\dagger l)^T C\, \tilde\phi^\dagger l                          
\end{equation}
The operator in (\ref{lvdim3}) is the equivalent of (\ref{lvdim5})
in a theory without a light Standard-Model Higgs boson.
In the limit $U\to 1$, (\ref{lvdim3}) reduces to $\nu^T_L C\nu_L$,
the Majorana mass term of the left-handed neutrinos.
Correspondingly, the very small neutrino masses set the scale for
the coefficient of $Q_{LV}$ in the Lagrangian. The smallness of this
coefficient can be understood in the standard way invoking the seesaw
mechanism. In this scenario right-handed neutrinos with large Majorana masses 
can be included into the effective Lagrangian by adding the term
\begin{equation}\label{leffnur}
{\cal L}_{\nu_R}=-M_D\, \bar l UP_+\eta-\frac{1}{2}M_R\, \nu^T_R C\nu_R
+ {\rm h.c.} 
\end{equation}
Integrating out $\nu_R$ leads to $Q_{LV}$ in (\ref{lvdim3}).
This consideration illustrates that a physical Higgs boson is not
essential for generating neutrino masses of either Dirac or Majorana
type.

Since the leading lepton-number violating operator in (\ref{lvdim3}) has
a very small coefficient, we will not list explicitly the formally
next-to-leading order operators in the $L$-violating sector.




\section{Standard Model with a heavy Higgs boson}
\label{sec:smhh}

In the following section we would like to illustrate how the nonlinear 
effective Lagrangian of electroweak interactions, including
next-to-leading order terms, arises in the context of a specific model. 
For this purpose we consider the
conventional Standard Model with a heavy Higgs boson of mass 
$m_h\sim\Lambda\gg v$, which is integrated out at scales below $m_h$. 
Although by itself 
the SM with a heavy Higgs particle is not a realistic description of Nature, 
it provides the simplest example of a renormalizable theory that
reduces to the nonlinear effective Lagrangian at electroweak energies.
A similar discussion in the context of chiral perturbation theory
and the linear $\sigma$-model, applied to low-energy QCD,
can be found for instance in \cite{Donoghue:1992dd}.

The Higgs sector of the renormalizable Standard Model can be
written as 
\begin{eqnarray}\label{lsmh}
{\cal L}_H &=& \frac{1}{4}\langle D_\mu H^\dagger D^\mu H\rangle +
\frac{\mu^2}{4}\langle H^\dagger H\rangle-
\frac{\lambda}{16}\langle H^\dagger H\rangle ^2 \nonumber\\
&& -\left( \bar qY_uH P_+r +                                      
\bar qY_dH P_-r +\bar lY_eH P_-\eta + {\rm h.c.}\right)
\end{eqnarray}
where the ordinary Higgs doublet, written in matrix notation
$H=(\tilde\phi, \phi)$, has been parametrized as
(see e.g. \cite{Isidori:2009ww})
\begin{equation}\label{hhu}
H\equiv (v+h) U\, ,\qquad U^\dagger U=1
\end{equation}
Here $U$ is the $SU(2)$ matrix field defined in (\ref{uudef}) and $h$ is the
Higgs boson, which transforms as a singlet under the SM gauge group.
With (\ref{hhu}) the Lagrangian in (\ref{lsmh}) becomes
\begin{equation}\label{lhuh}
{\cal L}_H=\frac{v^2}{4}\langle D_\mu U^\dagger D^\mu U\rangle-
v \left( \bar qY_u U P_+r +\ldots\right) + {\cal L}_{H,h}
\end{equation}
where
\begin{equation}\label{lhhj}
{\cal L}_{H,h}=
\frac{1}{2}h\left(-\partial^2 - m^2_h \right)h-
\frac{m^2_h}{2v} h^3-\frac{m^2_h}{8v^2}h^4+h J_1+\frac{h^2}{2}J_2
\end{equation}
Here we have used $m^2_h=2\mu^2$ and $\lambda=\mu^2/v^2$, and the
definitions
\begin{equation}\label{j12def}
J_1=\frac{v}{2}\langle D_\mu U^\dagger D^\mu U\rangle-
\left( \bar qY_u U P_+r +\ldots\right)\, ,\qquad
J_2=\frac{1}{2}\langle D_\mu U^\dagger D^\mu U\rangle 
\end{equation}
Removing the heavy Higgs boson $h$ from the theory, ${\cal L}_H$ in
(\ref{lhuh}) reduces to the lowest-order electroweak chiral Lagrangian.
Subleading terms in this Lagrangian are generated when corrections from
virtual $h$ are taken into account. Keeping only the effects of single
tree-level exchange of the heavy Higgs particle, the next-to-leading order
terms read
\begin{eqnarray}\label{lhheff}                                              
&& {\cal L}^{eff}_{H,h} = \frac{J^2_1}{2m^2_h}=\nonumber\\
&&\frac{v^2}{8 m^2_h}\langle D_\mu U^\dagger D^\mu U\rangle^2 -
\frac{v}{2m^2_h}\langle D_\mu U^\dagger D^\mu U\rangle\,
\left(\bar qY_uU P_+r +\bar qY_dU P_-r +\bar lY_eU P_-\eta + {\rm h.c.}\right)
\nonumber\\            
&&+\frac{1}{2m^2_h} \left(\bar qY_uU P_+r +\bar qY_dU P_-r +
\bar lY_eU P_-\eta + {\rm h.c.}\right)^2                  
\end{eqnarray}
The Lagrangian ${\cal L}^{eff}_{H,h}$ arises from the terms
$-m^2_h h^2/2+h J_1$ in (\ref{lhhj}) upon completing the square
when integrating out $h$ in the path integral.
Further terms generated by the exchange of $h$ bosons at tree level,
such as $J^2_1 J_2$ or $J^3_1$, require more than one $h$ propagator.
They are therefore suppressed by additional powers of $1/m^2_h$.

The next-to-leading operators in (\ref{lhheff}) are a subset of the 
complete basis listed in section \ref{sec:lsmnlo}. They illustrate
how operators of different canonical dimension emerge
at a given order of the effective Lagrangian. In particular, 
operators of dimension 4, 5 and 6 are present in (\ref{lhheff}). 
Specifically, the following terms out of the basis of NLO
operators appear: The pure $U$-field operator 
${\cal O}_{D1}$, the fermion bilinear operators
${\cal O}_{\psi S1}$, ${\cal O}_{\psi S2}$, ${\cal O}_{\psi S7}$, and their
hermitian conjugates, as well as all 4-fermion operators
coming from the square of the Yukawa terms. 
Up to hermitian conjugates, there are 12 such 4-fermion operators,
which are given by
\begin{eqnarray}\label{ysqrd14}
\bar qUP_+r\, \bar rP_+ U^\dagger q &=& 
-\frac{1}{4} {\cal O}_{LR1}-\frac{1}{2} {\cal O}_{LR10}  \nonumber \\
\bar qUP_-r\, \bar rP_- U^\dagger q &=&   
-\frac{1}{4} {\cal O}_{LR3}+\frac{1}{2} {\cal O}_{LR12} \nonumber \\
\bar lUP_- \eta\, \bar\eta P_- U^\dagger l &=&
-\frac{1}{4} {\cal O}_{LR8}+\frac{1}{2} {\cal O}_{LR17} \nonumber \\
\bar qUP_-r\, \bar\eta P_- U^\dagger l &=&
-\frac{1}{4} {\cal O}_{LR9}+\frac{1}{2} {\cal O}_{LR18}
\end{eqnarray}
together with 
\begin{equation}\label{st5fy10}
{\cal O}_{ST5},\, {\cal O}_{ST9},\,
{\cal O}_{FY1},\, {\cal O}_{FY3},\, {\cal O}_{FY5},\,
{\cal O}_{FY7},\, {\cal O}_{FY9},\, {\cal O}_{FY10}
\end{equation}

\section{Standard Model with a light Higgs boson}
\label{sec:smlh}

In the conventional Standard Model with a physical Higgs field
the next-to-leading order corrections to the renormalizable part
are given by dimension-6 operators in the effective Lagrangian
(up to the single dimension-5 operator that generates Majorana masses
for neutrinos given in (\ref{lvdim5})).
The list of these operators has been compiled by Buchm\"uller and Wyler
in \cite{Buchmuller:1985jz}. It has recently been revised and updated in
a new systematic study by Grzadkowski et al. \cite{Grzadkowski:2010es}.
In this section we show how the complete list of operators in
\cite{Grzadkowski:2010es} can be reproduced from the next-to-leading
operators in our effective theory. We have checked the contents of
\cite{Grzadkowski:2010es} and fully agree with the results of this paper.
 
The matrix $H=(\tilde\phi, \phi)$, where $\phi$ is the linearly 
transforming Higgs field, transforms in the same way as the Goldstone
field $U$. We may therefore replace $U\to H$ in the operators of the
nonlinear theory without affecting the required transformation properties,
gauge invariance in particular. However, introducing the linear field $H$
implies two important differences with respect to the nonlinear version
based on $U$. First, we no longer have the constraint $U^\dagger U=1$,
so that factors of $\l H^\dagger H\r$ can now appear as independent 
variables in the Lagrangian 
(note however that $\l H^\dagger H T_3\r = 0$). Second, since the
lowest-order Lagrangian becomes renormalizable with the transition to the
linear Higgs field, the new-physics scale $\Lambda$ may be taken to
decouple from the electroweak scale, $v/\Lambda\to 0$. Therefore, the
field $H$ enters the power counting with its canonical dimension
${\rm dim}\, H=1$. The ordering of terms in the effective Lagrangian
is then governed by dimensional counting alone.
For example, at lowest order, where ${\cal L}={\cal L}_4+{\cal L}_U$
from (\ref{lsm4}) and (\ref{lulo}), one has to substitute $U\to H/v$ and add
the invariants $\l H^\dagger H\r$ and $(\l H^\dagger H\r)^2$
to recover all terms up to dimension 4 of the renormalizable Standard
Model. Note that with the replacement $U\to H/v$ the custodial symmetry 
breaking term in (\ref{lub1}) becomes an operator of dimension 6.

Applying this reasoning to the next-to-leading-order terms based on the
nonlinear theory, we first have to include additional operators that
involve the invariant $\l H^\dagger H\r\sim\phi^\dagger\phi$.
These are easily enumerated at the dimension-6 level.
The first possibility is the (unique) operator with six factors of $\phi$,
$(\phi^\dagger\phi)^3$ (class $\phi^6$ in \cite{Grzadkowski:2010es}).
Next we have $(\phi^\dagger\phi)^2$ in combination with a dimension-2
operator, which can only be $D^2$. This gives the operator
$(\phi^\dagger\phi)\partial^2(\phi^\dagger\phi)$ 
(class $\phi^4D^2$ in \cite{Grzadkowski:2010es}),
which is unique up to total derivatives.
Finally, one may write down the operators with a single factor of
$\phi^\dagger\phi$, multiplied by any gauge invariant operator of
dimension 4. The candidates for the latter are just the terms in
the leading-order renormalizable Lagrangian. This gives immediately the
three operators
\begin{equation}\label{psi2phi3}
\phi^\dagger\phi\, \bar le\phi,\qquad\quad
\phi^\dagger\phi\, \bar qd\phi,\qquad\quad
\phi^\dagger\phi\, \bar qu\tilde\phi,
\end{equation}
corresponding to the Yukawa terms 
(class $\psi^2\phi^3$ in \cite{Grzadkowski:2010es}).
Similarly one has $\phi^\dagger\phi\, X_{\mu\nu}X^{\mu\nu}$ and
$\phi^\dagger\phi\, X_{\mu\nu}\tilde X^{\mu\nu}$ for the gauge fields
$X=G$, $W$, $B$ (class $X^2\phi^2$ in \cite{Grzadkowski:2010es}). 
No new operators arise from the
Higgs-potential terms $m^2\phi^\dagger\phi$ and $(\phi^\dagger\phi)^2$,
nor from the fermion kinetic terms $\bar\psi i\!\not\!\!\! D\psi$, 
where the leading-order equations of motion can be applied to reduce
$\phi^\dagger\phi\,\bar\psi i\!\not\!\! D\psi$ to structures already
listed above. For the only remaining possibility we have, using equations 
of motion and integration by parts, \cite{Grzadkowski:2010es}
\begin{equation}\label{phikindim6}
\phi^\dagger\phi\, (D_\mu\phi)^\dagger D^\mu\phi=
\frac{1}{2}\phi^\dagger\phi\, \partial^2 (\phi^\dagger\phi) +\ldots
\end{equation} 
The operator on the right is the term in class $\phi^4D^2$ mentioned before  
and the ellipsis denotes total derivatives and further operators already
encountered above. If the Higgs boson kinetic term is written differently,
we would have $\phi^\dagger\phi\, \phi^\dagger D^2\phi$, which is again not
independent upon using the equations of motion.
In total, therefore, one finds 11 operators involving factors of 
$\phi^\dagger\phi$.

Next, there are the operators of dimension 6 that do not include any 
factor of $U$ and which are thus identical in the nonlinear and in
the linear theory. These are the 4-fermion operators without $U$ fields
($\psi^4$) and the operators with three factors of field-strength tensors
($X^3$).

A further class of operators are the terms in the nonlinear theory
that become operators of dimension larger than 6 upon replacing
$U\to H$. They are therefore absent in the linear Higgs theory at order
$1/\Lambda^2$. The operators of this type are those in our classes
$UD^4$, $\psi^2 U D^2$ and $\psi^4 U$.

This leaves the cases $X^2U$, $XUD^2$, $\psi^2UD$ and $\psi^2 UX$
of next-to-leading operators in the nonlinear theory, as well
as the operator in (\ref{lub1}).
They translate into dimension-6 operators of the Standard Model when 
$U\to H=(\tilde\phi, \phi)$, as described in the following.

The custodial-symmetry breaking operator in (\ref{lub1}) becomes,
up to a total divergence, 
\begin{equation}\label{uhp2}
(\l U^\dagger D_\mu UT_3\r )^2 \to 
-\frac{1}{4} (\phi^\dagger\phi)\, \partial^2 (\phi^\dagger\phi)
-(D_\mu\phi^\dagger\phi)\, (\phi^\dagger D^\mu\phi)
\end{equation}
The first term on the right has already been listed above.
The second term is the remaining operator of class $\phi^4D^2$
in \cite{Grzadkowski:2010es}.

Inspection of (\ref{xudev}) and (\ref{xudod}) shows that
most of the operators in our classes $X^2U$ and $XUD^2$ translate
into operators of dimension higher than 6. The exceptions are
\begin{equation}\label{uhxud}
B_{\mu\nu}\ \l U^\dagger W^{\mu\nu} U T_3\r \to
-B_{\mu\nu}\ \phi^\dagger W^{\mu\nu}\phi
\end{equation}
and the analogous term where $W_{\mu\nu}\to\tilde W_{\mu\nu}$. These two 
operators complete class $X^2\phi^2$ in \cite{Grzadkowski:2010es}. 

Among the operators of class $\psi^2 UD$ in (\ref{psi2d}),
some contain four factors of $U$-fields and become operators of
dimension 8 when we replace $U\to H$. 
The remaining structures reduce to the eight dimension-6 operators
in class $\psi^2\phi^2 D$ of \cite{Grzadkowski:2010es}:
\begin{equation}
\bar\psi\gamma^\mu \psi\ \l U^\dagger iD_\mu UT_3\r \to
-\frac{1}{2}\, \bar\psi\gamma^\mu \psi\, (\phi^\dagger iD_\mu\phi
-iD_\mu\phi^\dagger \phi),
\qquad\psi=q,\, u,\, d,\, l,\, e \nonumber
\end{equation}
\begin{equation}\label{uhpsi2d}
\bar\psi\gamma^\mu iD_\mu U U^\dagger \psi \to
2\, \bar\psi\gamma^\mu T^a \psi\, (\phi^\dagger T^a iD_\mu\phi
-iD_\mu\phi^\dagger T^a\phi), \qquad \psi=q,\, l
\end{equation}
\vspace*{0.001cm}
\begin{equation}
\bar u\gamma^\mu d\ \l U^\dagger iD_\mu UP_{21}\r  \to
\bar u\gamma^\mu d\ i\tilde\phi^\dagger D_\mu \phi \nonumber
\end{equation}
Finally, since
\begin{equation}\label{uhpsi2x}
UP_+ r\to\tilde\phi u,\qquad
UP_- r\to\phi d, \qquad UP_- \eta\to\phi e
\end{equation}
the eight operators in our class $\psi^2 UX$ (\ref{psi2x}) reduce
to the operators in class $\psi^2 X\phi$ of \cite{Grzadkowski:2010es}.
This completes the demonstration of how the full basis of
dimension-6 operators given in \cite{Grzadkowski:2010es} can be
recovered from the next-to-leading order terms in the nonlinear
Higgs model.

\section{Example of renormalization at NLO}
\label{sec:renorm}

A comprehensive discussion of renormalization 
of the effective theory is beyond the scope of the present 
paper. We limit ourselves to a particular example, which
should serve to illustrate the essential features
of renormalization at next-to-leading order in the electroweak
chiral Lagrangian. 

Consider the amplitude for $t_R t_R\to t_L t_L$ scattering, 
taking the gaugeless limit for simplicity.
The leading-order (LO) amplitude is given by the tree-level contributions 
from the LO Lagrangian in (\ref{lsm4}) and
(\ref{lulo}). It arises from the exchange of a neutral Goldstone boson 
and reads
\begin{equation}\label{a0tu}
A_0=i y^2\left( \frac{\tau_t}{t}-\frac{\tau_u}{u}\right)
\end{equation}
where $y$ is the top-quark Yukawa coupling, $t=(p_1-p_3)^2$,
$u=(p_1-p_4)^2$ are Mandelstam variables and the fermion spinors
are denoted by
\begin{equation}\label{tautu}
\tau_t\equiv \bar t_L(p_3)t_R(p_1)\, \bar t_L(p_4)t_R(p_2),\quad
\tau_u\equiv \bar t_L(p_4)t_R(p_1)\, \bar t_L(p_3)t_R(p_2)
\end{equation}
The NLO amplitude arises both from one-loop corrections based on the LO 
Lagrangian, and from tree-level contributions of the NLO Lagrangian. 
The Goldstone-boson loop diagram shown in Fig. \ref{fig:pcexmp}
gives the correction
\begin{equation}\label{a1loop}
A_{1,loop}=\frac{i}{16\pi^2} \frac{y^2}{v^2}\frac{3}{2}\left[ 
\tau_t \left(\ln\frac{\mu^2}{-t}+\frac{1}{\epsilon}-\gamma+\ln 4\pi +2\right)
- (t\to u) \right]
\end{equation}
This contribution is divergent. It needs to be renormalized
by the appropriate coun\-ter\-term, which is provided by the
following piece of the NLO Lagrangian (see eq. (\ref{psi4y17}))
\begin{equation}\label{lfy1}
\Delta {\cal L}=\frac{c}{\Lambda^2}\, {\cal O}_{FY1} 
\to \frac{c}{16\pi^2 v^2}\, \bar t_L t_R \bar t_L t_R
\end{equation}
where the last term gives the part of the operator that contributes
to the considered amplitude at tree level. This contribution is
 \begin{equation}\label{a1tree}
A_{1,tree}=\frac{2i c^{(0)}}{16\pi^2 v^2} \left(\tau_t - \tau_u \right)
\end{equation}
with the bare coefficient $c^{(0)}$. Absorbing the divergence in (\ref{a1loop})
into this coefficient by minimal subtraction (and dropping a constant)
one finds for the renormalized NLO contribution
\begin{equation}\label{a1tot}
A_1=A_{1,loop}+A_{1,tree}=\frac{i}{16\pi^2} \frac{y^2}{v^2}\frac{3}{2}\left[ 
\tau_t \left(\ln\frac{\mu^2}{-t}+\frac{4}{3 y^2} c(\mu)\right)
- (t\to u) \right]
\end{equation}
The NLO term $A_1$ is suppressed with respect to the LO amplitude $A_0$
as $p^2/\Lambda^2$, where $p\sim v$ is the typical momentum scale of 
the process. $A_1$ depends on $c(\mu)$, a free parameter of the effective
theory, which cancels the dependence on the arbitrary scale $\mu$.
Independently of the local term provided by $c$, $A_1$ reproduces
the correct logarithmic dependence of the amplitude on $t$ and $u$
at NLO. 

It is interesting to compare the effective-theory calculation
just described with the analogous calculation in the usual
Standard Model that includes a heavy Higgs boson. The latter model
can be viewed as the simplest UV completion of the electroweak
chiral Lagrangian.
The LO amplitude for $t_Rt_R\to t_Lt_L$ in the gaugeless limit
coincides with the result in the effective theory given in (\ref{a0tu}). 
At NLO, that is, at order $p^2/\Lambda^2$, the amplitude receives
a contribution from tree-level Higgs exchange
 \begin{equation}\label{ah1tree}
A^h_{1,tree}=i \frac{y^2}{m^2_h} (\tau_t - \tau_u )
\end{equation}
where the Higgs-boson mass $m_h$ is assumed to be large, 
$p^2\ll m^2_h  \lsim \Lambda^2=16\pi^2 v^2$.
Including 1-loop corrections, one obtains the following
term with (large) logarithms of ${\cal O}(\ln m^2_h/p^2)$
\begin{equation}\label{ah1loop}
A^h_{1,loop}=\frac{i}{16\pi^2} \frac{y^2}{v^2}\frac{3}{2}\left[ 
\tau_t \ln\frac{m^2_h}{-t} - (t\to u) \right]
\end{equation}
This term comes from the Higgs-exchange diagram with a Goldstone-boson 
loop as a self-energy insertion in the Higgs propagator. Other loop 
corrections do not give rise to a large logarithm and
in (\ref{ah1loop}) we have neglected all such nonlogarithmic terms.
Combining (\ref{ah1tree}) and (\ref{ah1loop}), the NLO correction
in the heavy-Higgs model reads
\begin{equation}\label{ah1tot}
A^h_1=A^h_{1,tree}+A^h_{1,loop}=
\frac{i}{16\pi^2} \frac{y^2}{v^2}\frac{3}{2}\left[ 
\tau_t \left(\ln\frac{m^2_h}{-t} +\frac{2}{3}\frac{16\pi^2 v^2}{m^2_h}\right) 
- (t\to u) \right]
\end{equation}
If the Higgs boson is moderately heavy, $p^2\ll m^2_h  \ll \Lambda^2$,
the Higgs sector is still perturbative. In this case, the tree-level
term $\sim v^2/m^2_h$ in (\ref{ah1tot}) is dominating over the
logarithmic piece. As $m_h$ increases towards $\Lambda$, the Higgs sector
becomes strongly coupled and (\ref{ah1tot}) ceases to be
reliable. Besides, the tree-level term starts being comparable to order-one
contributions we have neglected in (\ref{ah1tot}). In this situation
we might still consider (\ref{ah1tot}) as a rough estimate
of the full amplitude.
In any case, (\ref{ah1tot}) defines a particular UV completion
of the effective-theory amplitude (\ref{a1tot}). Within such a UV
completion, the coefficient $c$ in (\ref{a1tot}) can be determined. In the 
present example one finds, equating (\ref{a1tot}) and (\ref{ah1tot}), 
\begin{equation}\label{cmumh}
c(\mu)=
y^2\left(\frac{3}{4}\ln\frac{m^2_h}{\mu^2}+\frac{8\pi^2 v^2}{m^2_h}\right)
\end{equation}
The logarithmic piece can be fixed from the one-loop UV divergence of
the effective theory, without knowing the details of the
physics at the high-energy scale $\Lambda$. However, as the example
illustrates especially for a moderate $m_h$, the nonlogarithmic,
constant term may be sizable. In this case the leading logarithmic 
approximation will not be accurate. Similar considerations have been given
in \cite{Longhitano:1980tm}.

\section{Conclusions}
\label{sec:concl}

In this article we have studied an effective field theory (EFT) description of 
the Standard Model assuming that the mechanism for spontaneous electroweak 
symmetry breaking is triggered by strong dynamics at the TeV scale. 
The resulting theory contains the fields of the Standard Model but, instead 
of describing electroweak symmetry breaking by introducing a Higgs doublet 
(linear sigma model), one uses a nonlinear realization for the Goldstone 
modes.

This article focusses mainly on clarifying the systematics of such an 
approach: in particular, we discuss in detail the power-counting of the 
$v/\Lambda$ expansion, providing the formulae needed to determine the 
complete list of NLO operators. This list is presented in Landau gauge and, 
for phenomenological convenience, also in unitary gauge. At NLO there are 
11 operators without fermion fields, 25 operators involving fermion bilinears 
and 64 four-quark operators. It is worth recalling that not all of the 
previous operators are required as counterterms of our effective field theory. 
While in some cases it is 
relatively easy to identify them, a general and systematic identification of 
the counterterms would require to carry out the full renormalization program 
for the LO operators, something that lies beyond the scope of the 
present article.

Several checks have been performed to eliminate redundancies between 
operators. In particular, we have emphasized relations between 
operators from the use of integration by parts and the equations of 
motion \cite{Nyffeler:1999ap,Grojean:2006nn}. These relations affect the 
operators contributing to triple-gauge-boson interactions. 

As an additional cross-check we have reexpressed our basis of operators in 
the linear representation. This particular case corresponds to deriving the 
NLO (dimension-6) operators in the Standard Model with a Higgs boson. 
Our results independently confirm the recent analysis of 
\cite{Grzadkowski:2010es}. The sets of NLO operators in our scenario and 
in the presence of a Higgs boson are seen to be related, but there are 
systematic differences.
This is an additional motivation to study the phenomenology 
associated with our EFT basis of operators, especially in 
electroweak precision observables but also in top-quark physics.

Finally, it is worth stressing that the approach undertaken in this article 
assumes the minimal content of Standard Model fields: the electroweak 
symmetry breaking sector is characterized only by three Goldstone bosons, 
which are the longitudinal modes of the $W$ and $Z$ gauge bosons. Therefore, 
no extraneous particles are introduced. However,
further light, electroweak-scale particles might in principle exist.
It would be interesting to study the 
systematics of a similar EFT scenario under this generalization.

\appendix

\section{Derivation of the power-counting formula}
\label{sec:apppc}

To obtain the power-counting formula in (\ref{pcupsix}) we start from 
the different types of vertices contained in
the leading order Lagrangian (\ref{lsm4}) and (\ref{lulo}). 
Denoting left-handed (right-handed) fermions, Goldstone bosons and
gauge fields by $\psi_L$ ($\psi_R$), $\varphi$ and $X_\mu$, respectively,
these vertices have the schematic form of 
$\varphi^{2i}$  ($2i$ Goldstone boson interaction),
$\bar\psi_{L(R)}\psi_{R(L)}\varphi^k$ (Yukawa interaction),  
$X_\mu \varphi^l$, $X^2_\mu  \varphi^s$ (gauge-Goldstone boson interaction),
$X^4_\mu$, $X^3_\mu$ (gauge boson self interaction), and
$\bar\psi_{L(R)}\psi_{L(R)}X_\mu$ (fermion-gauge boson coupling).
For each vertex, the factors of momentum $p$, Yukawa coupling $y$,
gauge coupling $g$ and electroweak scale $v$, contributing to the power 
counting in any diagram, are given by:
\begin{equation}\label{pcvert}
\begin{array}{c|c|c|c|c|c|c}
\varphi^{2i} & \bar\psi_{L(R)}\psi_{R(L)}\varphi^k & X_\mu \varphi^l &
X^2_\mu  \varphi^s & \quad X^4_\mu\quad & \quad X^3_\mu\quad & 
\bar\psi_{L(R)}\psi_{L(R)}X_\mu \\
\hline
\hline
p^2/v^{2i-2} & y/v^{k-1} & gp/v^{l-1} &
g^2/v^{s-2} & g^2 & gp & g \\
\hline
n_i & \nu_k & m_l & r_s & x & u & z_L (z_R) 
\end{array} 
\end{equation}
Here the last line defines the number of the corresponding
vertices in a given diagram.
A diagram ${\cal D}$ with $L$ loops, ${\cal F}_L$ (${\cal F}_R$), ${\cal B}$,
and ${\cal V}$, fermion, Goldstone, and gauge field propagators,
$F_L$ ($F_R$), $B$, and $V$ external fermion, Goldstone, and gauge field
lines, and with the numbers of vertices introduced in (\ref{pcvert}),
contains the factors ($z\equiv z_L+z_R$)  
\begin{eqnarray}\label{pcd0}
{\cal D}&\sim & \frac{v^{2L}}{\Lambda^{2L}}
\frac{y^{\sum_k\nu_k} g^{\sum_l m_l+\sum_s 2r_s + 2x+u+z}}{
v^{\sum_i (2i-2)n_i + \sum_k(k-1)\nu_k +\sum_l (l-2)m_l+\sum_s(s-2)r_s}}
\nonumber\\
&& \times p^{4L+\sum_i 2n_i+\sum_l m_l + u-2{\cal B}-2{\cal V}-{\cal F}_L
-{\cal F}_R -V}\
\bar\psi_L^{F^1_L} \psi_L^{F^2_L} \bar\psi_R^{F^1_R} \psi_R^{F^2_R}\
\varphi^B\ (X_{\mu\nu})^V
\end{eqnarray}
In writing (\ref{pcd0}) we have made explicit the external fermion,
Goldstone-boson and gauge-boson fields. We have written the gauge field
in terms of the field strength $X_{\mu\nu}$ (rather than the
gauge potential $X_\mu$) and correspondingly associated a factor of $p$
with each gauge field, $X_{\mu\nu}\sim p X_\mu$.
This is because we are interested in the power counting for 
gauge-invariant operators containing factors of field-strength tensors.
Operators with factors of $X_\mu$, but without associated momentum $p$,
are not gauge-invariant by themselves and are related to
operators with (gauge-covariant) derivatives.
The first factor in (\ref{pcd0}) is the loop factor
$1/(16\pi^2)^L$, where we have identified $1/(16\pi^2)=v^2/\Lambda^2$.
The remaining factors follow immediately from collecting all parts
of the diagram.

The expression (\ref{pcd0}) can be put into a more useful form
by employing the well-known topological identities for Feynman graphs
\begin{equation}\label{topifl}
F_L+ 2{\cal F}_L = \sum_k\nu_k +2 z_L\, ,\qquad
F_R+ 2{\cal F}_R = \sum_k\nu_k +2 z_R 
\end{equation}
\begin{equation}\label{topibb}
B+ 2{\cal B} = \sum_i 2i n_i +\sum_k k\nu_k +\sum_l l m_l +\sum_s s r_s 
\end{equation}
\begin{equation}\label{topivv}
V+ 2{\cal V} = \sum_l m_l +\sum_s 2r_s+4x+3u +z 
\end{equation}
\begin{equation}\label{topill}
L = {\cal F}_L+{\cal F}_R+{\cal B}+{\cal V}-
\sum_i n_i-\sum_k \nu_k-\sum_l m_l -\sum_s r_s -x-u-z+1
\end{equation}
Using these five equations to eliminate the five quantities
${\cal F}_L$, ${\cal F}_R$, ${\cal B}$, ${\cal V}$ and $L$, one finds
for the power $d$ of momentum factor $p$ in (\ref{pcd0})
\begin{equation}\label{naivedod}
d=4-B-2V-\frac{3}{2} F_L-\frac{3}{2} F_R+
\sum_i (2i-2)n_i + \sum_k (k-1)\nu_k +\sum_l (l-2)m_l+\sum_s(s-2)r_s
\end{equation} 
This corresponds to the standard expression for the superficial degree
of divergence of a diagram, where the nonrenormalizable interactions
are seen to give positive contributions that increase with the number
of vertices. This is a correct result, but still not immediately
useful for our purposes. To proceed, we may eliminate the positive
contributions of the nonrenormalizable terms with the help of
\begin{equation}\label{ll2bv}
2 L=2-B-V-F_L-F_R+
\sum_i (2i-2)n_i + \sum_k k\nu_k +\sum_l (l-1)m_l+\sum_s s r_s+2x+u+z
\end{equation}
which follows from (\ref{topill}) after eliminating 
${\cal F}_L$, ${\cal F}_R$, ${\cal B}$, ${\cal V}$.
Taking the difference between (\ref{naivedod}) and (\ref{ll2bv}),
the positive vertex terms in (\ref{naivedod}) cancel, and one has
\begin{equation}\label{dfinal}
d=2L+2-\frac{F_L+F_R}{2}-V-\sum_k \nu_k-\sum_l m_l-\sum_s 2r_s -2x-u-z  
\end{equation}
Using (\ref{dfinal}) and (\ref{ll2bv}) in (\ref{pcd0}) we finally
obtain (\ref{pcupsix}).

\section{Technical aspects of operator building}
\label{sec:appa}

\subsection{Operators without fermions}

For this subset of operators the most convenient method is to list the
elementary building blocks of $SU(2)$ algebra, as was done in
\cite{Longhitano:1980tm}. With $SU(2)$ elements, only traces of at
most 3 elements are not redundant. This follows from
\begin{equation}\label{ttdeleps}
T^a T^b=\frac{1}{4}\delta^{ab}+\frac{i}{2}\varepsilon^{abc} T^c\, ,
\end{equation}
which implies $\langle T^a T^b\rangle\sim \delta^{ab}$ and
$\langle T^a T^b T^c\rangle\sim \varepsilon^{abc}$. For our present 
purposes we will only need as $SU(2)$ elements the chiral vector,
$L_{\mu}=UD_{\mu}U^{\dagger}$, $W_{\mu\nu}$ and the scalar spurion
$\tau_L=UT_3U^{\dagger}$, all of which are (anti-) hermitean and traceless. 
That singles out the following elementary building
blocks:
\begin{align}\label{blocks}
&\langle L_{\mu}L_{\nu}\rangle,&\langle W_{\mu\nu}L_{\lambda}\rangle,\nonumber\\
&\langle \tau_LL_{\mu}\rangle,&\langle \tau_LW_{\mu\nu}\rangle,\nonumber\\
&\langle L_{\mu}L_{\nu}L_{\lambda}\rangle,&\langle W_{\mu\nu}L_{\lambda}L_{\rho}\rangle
,\nonumber\\
&\langle \tau_L L_{\mu} L_{\nu}\rangle,&\langle \tau_LW_{\mu\nu}L_{\lambda}\rangle.
\end{align}
The previous building blocks have to be assembled into the dimension-4 
operators of classes $UD^4$, $X^2U$ and $XUD^2$, such that traces of
3 elements only show up once in any operator (since the product of two
Levi-Civita tensors is reducible). Lorentz invariance then naturally selects
the possible combinations. It is convenient to separate the possible
operators into two categories: those with and without gauge field strengths.
To NLO the former set consists of
\begin{align}\label{pure}
&\langle L_{\mu}L_{\nu}\rangle\langle L^{\mu}L^{\nu}\rangle,
&\langle L_{\mu}L^{\mu}\rangle\langle L_{\nu}L^{\nu}\rangle,\nonumber\\
&\langle L_{\nu}L^{\nu}\rangle\langle \tau_LL_{\mu}\rangle^2,
&\langle L_{\mu}L_{\nu}\rangle\langle \tau_LL^{\mu}\rangle
\langle \tau_LL^{\nu}\rangle,\nonumber\\
&\langle \tau_LL_{\mu}\rangle^2\langle \tau_LL_{\nu}\rangle^2.
\end{align}
The list consists only of CP-conserving operators: operators involving
purely Goldstone fields are symmetric under the exchange of at least two
Lorentz indices and therefore cancel identically when contracted with the
Levi-Civita symbol $\varepsilon_{\mu\nu\lambda\rho}$.

In contrast, operators with gauge field strengths (at NLO) have contributions
from CP-conserving and CP-violating sectors. We will discuss the CP-conserving
sector in detail and later on generalize to the CP-violating sector. The set 
of CP-conserving
operators one can construct with the building blocks of Eq.~(\ref{blocks}) is
\begin{align}\label{naive1}
{\cal{O}}_{XU1}&=g^{\prime}gB_{\mu\nu}\langle W^{\mu\nu}\tau_L\rangle\nonumber\\
{\cal{O}}_{XU2}&=g^2 \langle W_{\mu\nu}\tau_L\rangle^2\nonumber\\
{\cal{O}}_{XU3}&=g\varepsilon_{\mu\nu\lambda\rho}
\langle W^{\mu\nu}L^{\lambda}\rangle\langle\tau_L L^{\rho}\rangle\nonumber\\
{\cal{O}}_{XU7}&=ig^{\prime}B_{\mu\nu}\langle\tau_L[L^{\mu},L^{\nu}]
\rangle\nonumber\\
{\cal{O}}_{XU8}&=ig\langle W_{\mu\nu}[L^{\mu},L^{\nu}]
\rangle\nonumber\\
{\cal{O}}_{XU9}&=ig\langle W_{\mu\nu}\tau_L\rangle 
\langle \tau_L[L^{\mu},L^{\nu}]\rangle
\end{align}
One further operator that can be built in this way is not independent:
\begin{equation}\label{taulw}
\langle [\tau_L,L^\mu]W_{\mu\nu}\rangle\langle\tau_L L^\nu\rangle=
\frac{1}{2}\langle\tau_L W_{\mu\nu}\rangle
\langle\tau_L [L^\mu,L^\nu]\rangle
-\frac{1}{4}\langle W_{\mu\nu} [L^\mu,L^\nu]\rangle
\end{equation}
To demonstrate this, we start from the identity
($\tau^a_L\equiv U T^a U^\dagger$)
\begin{align}
\langle \tau_L L_{\mu}W^{\mu\nu}\rangle\langle\tau_L L_{\nu}\rangle
&=\langle \tau_L^a L_{\mu}W^{\mu\nu}\rangle\langle\tau_L^a L_{\nu}\rangle - 
\langle \tau_L^i L_{\mu}W^{\mu\nu}\rangle\langle\tau_L^i L_{\nu}\rangle
\qquad (i\neq 3)
\end{align}
However,
\begin{align}
\langle \tau_L^i L_{\mu}W^{\mu\nu}\rangle\langle\tau_L^i L_{\nu}\rangle
&=\frac{i}{8}\varepsilon^{ijk}
(U^\dagger L_{\nu}U)^i (U^\dagger L_{\mu}U)^j (U^\dagger W^{\mu\nu}U)^k \nonumber\\
&=\frac{i}{8}\varepsilon^{i3k}
(U^\dagger L_{\nu}U)^i (U^\dagger L_{\mu}U)^3 (U^\dagger W^{\mu\nu}U)^k\nonumber\\
&\quad +\frac{i}{8}\varepsilon^{ij3}
(U^\dagger L_{\nu}U)^i (U^\dagger L_{\mu}U)^j (U^\dagger W^{\mu\nu}U)^3\nonumber\\
&=-\langle \tau_L L_{\nu}W^{\mu\nu}\rangle\langle L_{\mu}\tau_L\rangle+
\langle L_{\nu} L_{\mu} \tau_L \rangle\langle W^{\mu\nu}\tau_L\rangle
\end{align}
The terms with $\tau_L^a$ can be simplified with the $SU(2)$ relation 
$(\tau^a_L)_{ij}(\tau^a_L)_{kl}=
\frac{1}{2}\delta_{il}\delta_{jk}-\frac{1}{4}\delta_{ij}\delta_{kl}$ 
and one finally obtains (\ref{taulw}). 


Contrary to the list in (\ref{pure}), three of the operators in 
(\ref{naive1}) can be shown to be 
redundant \cite{Nyffeler:1999ap,Grojean:2006nn}. By using the identities
\begin{align}
D_{\mu}\tau_L&=[\tau_L,L_{\mu}],\label{first}\\
D_\mu L_\nu-D_\nu L_\mu
&=-igW_{\mu\nu}+ig^{\prime}B_{\mu\nu}\tau_L-[L_{\mu},L_{\nu}],
\label{second}
\end{align}
integration by parts and the equations of motion for the gauge fields,
\begin{align}
\partial^{\mu}B_{\mu\nu}&=g^{\prime}\left[Y_{\psi}{\bar{\psi}}\gamma_{\nu}\psi+
\frac{i}{2}v^2\langle\tau_LL_{\nu}\rangle\right]\nonumber\\
D^{\mu}W_{\mu\nu}^a&=g\left[{\bar{\psi}}_L\gamma_{\nu}T^a\psi_L
-\frac{i}{2}v^2\langle T^aL_{\nu}\rangle\right]
\end{align}
one finds that
\begin{align}
B_{\mu\nu}\langle\tau_L[L^{\mu},L^{\nu}]\rangle
&=-2g^{\prime}Y_{\psi}{\bar{\psi}}\gamma_{\mu}\psi\langle\tau_LL^{\mu}\rangle
-ig^{\prime}v^2\langle\tau_LL_{\mu}\rangle^2+igB^{\mu\nu}
\langle\tau_LW_{\mu\nu}\rangle-\frac{ig^{\prime}}{2}B_{\mu\nu}B^{\mu\nu}\nonumber\\
\langle W_{\mu\nu}[L^{\mu},L^{\nu}]\rangle
&=g{\bar{\psi}}_L\gamma_{\mu}L^{\mu}\psi_L
-\frac{ig}{2}v^2\langle L_{\mu}L^{\mu}\rangle+ig^{\prime}B^{\mu\nu}
\langle\tau_LW_{\mu\nu}\rangle-ig\langle W_{\mu\nu}W^{\mu\nu}\rangle\nonumber\\
\langle\tau_LW_{\mu\nu}\rangle\langle\tau_L[L^{\mu},L^{\nu}]\rangle&=
-\frac{g}{2}{\bar{\psi}}_L\gamma_{\mu}\tau_L\psi_L\langle\tau_LL^{\mu}\rangle
+\frac{ig}{4}v^2\langle\tau_LL_{\mu}\rangle^2
+\frac{ig}{2}\langle\tau_LW_{\mu\nu}\rangle^2\nonumber\\
&\quad +\frac{g}{4}\bar\psi_L\gamma_\mu L^\mu\psi_L-\frac{ig}{8} v^2
\langle L_\mu L^\mu\rangle-\frac{ig}{4}\langle W_{\mu\nu} W^{\mu\nu}\rangle
\end{align}
and thus ${\cal{O}}_{XU7}$ ${\cal{O}}_{XU8}$ and ${\cal{O}}_{XU9}$ above can 
be eliminated.

For the CP-violating sector one can likewise write:
\begin{align}\label{naive2}
{\cal{O}}_{XU4}
&=g^{\prime}g\varepsilon_{\mu\nu\lambda\rho}\langle \tau_L W^{\mu\nu}\rangle
B^{\lambda\rho}\nonumber\\
{\cal{O}}_{XU5}&=g^2\varepsilon_{\mu\nu\lambda\rho}\langle\tau_LW^{\mu\nu}\rangle
\langle\tau_LW^{\lambda\rho}\rangle\nonumber\\
{\cal{O}}_{XU6}&=g\langle W_{\mu\nu}L^{\mu}\rangle\langle \tau_LL^{\nu}\rangle
\nonumber\\
{\cal{O}}_{XU10}
&=ig^{\prime}\varepsilon_{\mu\nu\lambda\rho}B^{\mu\nu}\langle\tau_L[L^{\lambda},L^{\rho}]
\rangle\nonumber\\
{\cal{O}}_{XU11}
&=ig\varepsilon_{\mu\nu\lambda\rho}\langle W^{\mu\nu}[L^{\lambda},L^{\rho}]
\rangle\nonumber\\
{\cal{O}}_{XU12}&=ig\varepsilon_{\mu\nu\lambda\rho}\langle W^{\mu\nu}\tau_L\rangle 
\langle \tau_L[L^{\lambda},L^{\rho}]
\rangle
\end{align}
Similarly to the list in (\ref{naive1}), there is no further independent
structure beyond the six operators shown. This is because (\ref{taulw}) also
holds when $W_{\mu\nu}\to\tilde W_{\mu\nu}$. 
Proceeding in the same way as for the CP-conserving sector, and recalling 
the Bianchi identity 
$\varepsilon_{\mu\nu\lambda\rho}\partial^{\mu}B^{\nu\lambda}=0=
\varepsilon_{\mu\nu\lambda\rho}D^{\mu}W^{\nu\lambda}$, 
one can show the relations
\begin{align}
\varepsilon_{\mu\nu\lambda\rho}B^{\mu\nu}\langle\tau_L[L^{\lambda},L^{\rho}]\rangle
&=ig\varepsilon_{\mu\nu\lambda\rho}B^{\mu\nu}
\langle\tau_LW^{\lambda\rho}\rangle\nonumber\\
\varepsilon_{\mu\nu\lambda\rho}\langle W^{\mu\nu}[L^{\lambda},L^{\rho}]\rangle
&=ig^{\prime}\varepsilon_{\mu\nu\lambda\rho}B^{\mu\nu}
\langle\tau_LW^{\lambda\rho}\rangle
\nonumber\\
\varepsilon_{\mu\nu\lambda\rho}\langle\tau_LW^{\mu\nu}\rangle
\langle\tau_L[L^{\lambda},L^{\rho}]\rangle
&=\frac{ig}{2}\varepsilon_{\mu\nu\lambda\rho}
\langle\tau_LW^{\mu\nu}\rangle\langle\tau_LW^{\lambda\rho}\rangle
\end{align}
and therefore ${\cal{O}}_{XU10}$ ${\cal{O}}_{XU11}$ and ${\cal{O}}_{XU12}$ 
above are redundant. The previous relations hold up to total derivatives, 
{\it{i.e.}}, also neglecting the anomalous operator 
${\tilde{W}}_{\mu\nu}W^{\mu\nu}$.


\subsection{Four-fermion operators}

The $({\bar{L}}L)({\bar{L}}L)$, $({\bar{R}}R)({\bar{R}}R)$ and 
$({\bar{L}}L)({\bar{R}}R)$ groups are rather straightforward to dress with 
chiral fields. In order to study the groups $({\bar{L}}R)({\bar{L}}R)$ and 
$({\bar{L}}R)({\bar{R}}L)$ it is convenient to parametrize the chiral 
field as follows:
\begin{align}
U=\left(
\begin{array}{cc}
e^{i\alpha}\cos\theta & -e^{i\beta}\sin\theta\\
e^{-i\beta}\sin\theta & e^{-i\alpha}\cos\theta
\end{array}
\right)\equiv (\tilde{\omega},\omega)
\end{align}
The $SU(2)$ doublets trivially satisfy 
$\omega^{\dagger}\omega=\tilde{\omega}^{\dagger}\tilde\omega=1$. The unitarity 
of $U$ also implies 
$\tilde\omega\tilde{\omega}^\dagger+\omega\omega^\dagger={\mathbf{1}}$. 

One can show that (lepton and baryon-conserving) 4-fermion operators can 
only appear as hypercharge $Y=0,\pm1$. This follows from an explicit 
enumeration of all possibilities. For the first category the 
4-fermion operators have to be dressed with $\omega,\tilde{\omega}$, to 
ensure hypercharge conservation. The following projections turn out to be 
useful  
\begin{align}
UP_+&=(\tilde{\omega},0), & UP_{12}&=(0,\tilde{\omega})\nonumber\\
UP_{21}&=(\omega,0), & UP_-&=(0,\omega)
\end{align}
For instance, for the combination $({\bar{l}}e_R)({\bar{q}}u_R)$, one finds 
the two possibilites:
\begin{align}
({\bar{l}}\tilde\omega e_R)~({\bar{q}}\omega u_R)
&=({\bar{l}}UP_{12}\eta)~({\bar{q}}UP_{21}r)\nonumber\\
({\bar{l}}\omega e_R)~({\bar{q}}\tilde\omega u_R)
&=({\bar{l}}UP_-\eta)~({\bar{q}}UP_+r)
\end{align}
For the $Y=\pm1$ 4-fermion operators, one achieves global neutral 
hypercharge dressing with the combinations $\omega,\omega$ and $\tilde\omega,
\tilde\omega$. However, notice that the resulting operators are not 
independent, but rather related by hermitian conjugation.
 

\subsection{Fermion bilinear operators}

For this sector it is more convenient to work with the generic four 
combinations (signs are uncorrelated):
\begin{align}\label{proj}
P_{\pm} {\cal{A}} P_{\pm},
\end{align}
where $P_{\pm}=\frac{1}{2}\pm T_3$ are flavor projectors and ${\cal{A}}$ is 
any traceless, hermitean $SU(2)$ matrix 
(for instance $iL_{\mu}$, $\tau_L$, or $W_{\mu\nu}$). 
As such, ${\cal{A}}$ has only two independent elements and can be 
parametrized as
\begin{equation}
{\cal{A}}=\left(
\begin{array}{cc}
a & b\\
b^{\dagger} & -a
\end{array}
\right).
\end{equation}
Notice that Eq.~(\ref{proj}) actually selects the different components of 
${\cal{A}}$, namely
\begin{align}
P_+ {\cal{A}}P_+&=aP_+,\nonumber\\
P_+ {\cal{A}}P_-&=bP_{12},\nonumber\\
P_- {\cal{A}}P_+&=b^{\dagger}P_{21},\nonumber\\
P_- {\cal{A}}P_-&=-aP_-,
\end{align}
where 
\begin{align}
a&={\mathrm{tr}}(T_3 {\cal{A}}),\nonumber\\
b&={\mathrm{tr}}(P_{21} {\cal{A}}),\nonumber\\
b^{\dagger}&={\mathrm{tr}}(P_{12} {\cal{A}}),
\end{align}
and $P_{ij}$ are the $SU(2)$ ladder operators. Inserting as many times 
Eq.~(\ref{proj}) between fermion bilinears as needed, one can easily exhaust 
the list of operators for any dimension. Notice that the products of 
projectors have to be conveniently threaded using $P_+P_+=P_+$, $P_-P_-=P_-$ 
and $P_+P_-=P_-P_+=0$. For instance, consider NLO operators to the Yukawa 
terms $\bar Q_LUP_{\pm}Q_R$. One has to consider operators stemming from
\begin{align}
{\bar{Q}}_L U[P_{\pm}R_{\mu}P_{\pm}R^{\mu}P_{\pm}]Q_R,
\end{align}
where $R_{\mu}=U^{\dagger}D_{\mu}U$. Out of the 8 possibilities to connect the 
two projectors, only 6 are independent (up to $h.c.$), namely
\begin{equation}
{\bar{Q}}_{L}UP_{\pm}Q_R\langle\tau_LL_{\mu}\rangle^2,\qquad
{\bar{Q}}_{L}UP_{\pm}Q_R\langle L^{\mu}L_{\mu}\rangle,\nonumber
\end{equation}
\begin{equation}
\bar Q_L UP_{12}Q_R\langle P_{21}R_\mu\rangle\langle\tau_L L^\mu\rangle,
\qquad \bar Q_L UP_{21}Q_R\langle P_{12}R_\mu\rangle\langle\tau_L L^\mu\rangle.
\end{equation}
Analogously, NLO operators to the 
left-handed vector current will come from
\begin{equation} 
\bar Q_L\gamma^{\mu}U[P_{\pm}R_{\mu}P_{\pm}]U^{\dagger}Q_L .
\end{equation} 
Out of the 4 possibilities, 3 are linearly independent (up to $h.c.$):
\begin{equation}
i{\bar{Q}}_{L}\gamma_{\mu}Q_{L}\langle\tau_LL^{\mu}\rangle,\quad
i{\bar{Q}}_{L}\gamma_{\mu}\tau_LQ_{L}\langle\tau_LL^{\mu}\rangle,\quad
i\bar Q_L\gamma_\mu UP_{12} U^\dagger Q_L \langle P_{21}R^\mu\rangle.
\end{equation}

\section{\boldmath  NLO operators in unitary gauge}
\label{sec:appb}

In order to get a clearer picture of the phenomenological 
consequences of the NLO operators, it is convenient to write 
them down in unitary gauge. In this gauge only physical particles are 
present, that is, $U=1$ and the Goldstone bosons are traded in for the 
longitudinal modes of the $W$ and $Z$ gauge bosons. In the following, we will 
rotate the $W_{\mu}^{a}$ and $B_{\mu}$ fields and express operators in 
terms of physical gauge bosons. An exception will be made with the 
operators ${\cal{O}}_{XUj}$, which are more conveniently expressed  
in terms of $W_{\mu\nu}^{3}$ and $B_{\mu}$, 
since they contribute to the oblique parameters $S$, $T$ and $U$.

The leading order Lagrangian ${\cal{L}}_U$ then reduces to the mass term 
for the gauge bosons and fermions:
\begin{align}
{\cal L}_U&=\frac{v^2}{8}\left[2g^2 W_\mu^+ W^{\mu-}+(g^2+g^{\prime 2}) 
Z_{\mu}Z^{\mu}\right]-v \left( \bar u_L Y_u u_R + 
\bar d_L Y_d d_R +\bar e_L Y_e e_R + {\rm h.c.}\right)
\end{align}
The leading custodial symmetry breaking operator (\ref{lub1}) can be 
expressed as
\begin{equation}\label{lub1u}
{\cal L}_{\beta_1}=-\frac{\beta_1v^2}{4} \left(gW_{\mu}^{3}-g^{\prime}B_{\mu} 
\right)^2 \equiv -\frac{\beta_1v^2}{4}(g^2+g'^2)Z_{\mu}Z^{\mu}
\end{equation}
Likewise, one can list the NLO operators for each class 
of section~\ref{sec:lsmnlo} (in the following 
we will omit those operators that in a general gauge do not contain the 
chiral field $U$): 
\begin{align}  
{\cal O}_{D1}&=\frac{1}{4}\left[2g^2 W_\mu^+ W^{\mu-}+(g^2+g^{\prime 2}) 
Z_{\mu}Z^{\mu}\right]^2\nonumber\\
{\cal O}_{D2}&=\frac{1}{4} \left[ g^2 (W_\mu^+ W^-_\nu + W_\nu^+ W^-_\mu) +
(g^2+g^{\prime 2}) Z_{\mu}Z_{\nu}\right]^2\nonumber\\
{\cal O}_{D3}&=\frac{1}{16}\left[(g^2+g'^2)Z_{\mu}Z^{\mu}\right]^2\nonumber\\
{\cal O}_{D4}&=-\frac{1}{8}\left[(g^2+g'^2)Z_{\mu}Z^{\mu}\right]
\left[2g^2 W_\nu^+ W^{-\nu}+(g^2+g^{\prime 2}) Z_{\nu}Z^{\nu}\right]\nonumber\\
{\cal O}_{D5}&= -\frac{1}{8}\left[(g^2+g'^2)Z^{\mu}Z^{\nu}\right]
\left[2 g^2 W_\mu^+ W_\nu^-  + (g^2+g^{\prime 2}) Z_{\mu}Z_{\nu}\right]\nonumber
\end{align}
\begin{center}
\rule{5.5cm}{0.01cm}
\end{center}
\begin{align}
{\cal O}_{XU1} &=\frac{g'g}{2}\ B^{\mu\nu}W_{\mu\nu}^{3} &
{\cal O}_{XU2} &=\frac{g^2}{4}W_{\mu\nu}^{3}W^{3\mu\nu}\nonumber\\
{\cal O}_{XU3}&=\frac{g}{4}\varepsilon^{\mu\nu\lambda\rho}
\left[gW_{\mu\nu}^{a}W_{\lambda}^{a}-g'W_{\mu\nu}^{3}B_{\lambda}\right]
\left[g'B_{\rho}-gW_{\rho}^{3}\right] & \nonumber\\
{\cal O}_{XU4} &= \frac{g'g}{2}\varepsilon^{\mu\nu\lambda\rho}
B_{\mu\nu}W_{\lambda\rho}^{3} &          
{\cal O}_{XU5} &= \frac{g^2}{4}\varepsilon^{\mu\nu\lambda\rho} 
W_{\mu\nu}^{3}W_{\lambda\rho}^{3}\nonumber\\  
{\cal O}_{XU6} &= \frac{g}{4} \left[gW_{\mu\nu}^{a}W^{a\mu}-
g' W_{\mu\nu}^{3}B^{\mu}\right]\left[g' B^{\nu} - g W^{3\nu} \right]& \nonumber
\end{align}
\begin{center}
\rule{5.5cm}{0.01cm}
\end{center}
\begin{align}
{\cal O}_{\psi V1}&=-\displaystyle\frac{\sqrt{g^2+g'^2}}{2}\bar q\gamma^\mu q 
Z_{\mu} &
{\cal O}_{\psi V2}&=-\displaystyle\frac{\sqrt{g^2+g'^2}}{2}\bar q\gamma^\mu T_3 q 
Z_{\mu}\nonumber\\
{\cal O}_{\psi V3}&=-\displaystyle\frac{g}{\sqrt{2}}\bar u_L\gamma^\mu d_L 
 W_{\mu}^+  &
{\cal O}_{\psi V4}&=-\displaystyle\frac{\sqrt{g^2+g'^2}}{2}\bar u_R\gamma^\mu u_R 
Z_{\mu}\nonumber\\
{\cal O}_{\psi V5}&=-\displaystyle\frac{\sqrt{g^2+g'^2}}{2}\bar d_R\gamma^\mu d_R 
Z_{\mu} &
{\cal O}_{\psi V6}&=-\displaystyle\frac{g}{\sqrt{2}}\bar u_R\gamma^\mu d_R 
 W_{\mu}^+ \nonumber\\
{\cal O}_{\psi V7}&=-\displaystyle\frac{\sqrt{g^2+g'^2}}{2}\bar l\gamma^\mu l 
Z_{\mu} &
{\cal O}_{\psi V8}&=-\displaystyle\frac{\sqrt{g^2+g'^2}}{2}\bar l\gamma^\mu T_3 l 
Z_{\mu}\nonumber\\
{\cal O}_{\psi V9}&=-\displaystyle\frac{g}{\sqrt{2}} \bar\nu_L \gamma^{\mu} e_L 
 W_{\mu}^+  &
{\cal O}_{\psi V10}&=
-\displaystyle\frac{\sqrt{g^2+g'^2}}{2}\bar e_R\gamma^\mu e_R  Z_{\mu}\nonumber
\end{align}
\begin{center}
\rule{5.5cm}{0.01cm}
\end{center}
\begin{align}
{\cal O}_{\psi S1,2} &=\frac{1}{2}({\bar{u}}_Lu_R,{\bar{d}}_Ld_R)\!\! 
\left[2g^2 W_\mu^+ W^{-\mu}+(g^2+g^{\prime 2}) Z_{\mu}Z^{\mu}\right] 
\nonumber\\
{\cal O}_{\psi S3,4} &=- \frac{g^2+g'^2}{4} ({\bar{u}}_Lu_R,{\bar{d}}_Ld_R) 
Z_{\mu}Z^{\mu} \nonumber\\
{\cal O}_{\psi S5} &= -\frac{g\sqrt{g^2+g'^2}}{2\sqrt{2}}{\bar{u}}_Ld_R 
W_{\mu}^+Z^{\mu} \qquad
{\cal O}_{\psi S6} = -\frac{g\sqrt{g^2+g'^2}}{2\sqrt{2}}{\bar{d}}_Lu_R  
W_{\mu}^-Z^{\mu} \nonumber\\
{\cal O}_{\psi S7} &= \frac{1}{2}{\bar{e}}_Le_R \left[2g^2 W_\mu^+ W^{-\mu}+
(g^2+g^{\prime 2}) Z_{\mu}Z^{\mu}\right] \nonumber\\
{\cal O}_{\psi S8} &= -\frac{g^2+g'^2}{4} {\bar{e}}_Le_R Z_{\mu}Z^{\mu} \qquad
{\cal O}_{\psi S9} = -\frac{g\sqrt{g^2+g'^2}}{2\sqrt{2}}{\bar{\nu}}_Le_R 
W_{\mu}^+Z^{\mu}\nonumber                     
\end{align}
\begin{center}
\rule{5.5cm}{0.01cm}
\end{center}
\begin{align}
{\cal O}_{\psi T1} &= -\frac{g\sqrt{g^2+g'^2}}{2\sqrt{2}}
\bar u_L\sigma^{\mu\nu} d_R  W_{\mu}^+Z_{\nu} &  
{\cal O}_{\psi T2} &= -\frac{g\sqrt{g^2+g'^2}}{2\sqrt{2}}
\bar d_L\sigma^{\mu\nu} u_R W_{\mu}^-Z_{\nu} \nonumber\\    
{\cal O}_{\psi T3,4} &=\frac{g^2}{2} ({\bar{u}}_L\sigma^{\mu\nu}u_R,
{\bar{d}}_L\sigma^{\mu\nu}d_R)W^+_\mu W^-_\nu  & \nonumber\\  
{\cal O}_{\psi T5} &= 
-\frac{g\sqrt{g^2+g'^2}}{2\sqrt{2}}\bar{\nu}_L \sigma^{\mu\nu} e_R 
W_{\mu}^+Z_{\nu} &      
{\cal O}_{\psi T6} &= \frac{g^2}{2} \bar e_L \sigma^{\mu\nu} e_R 
W^+_\mu W^-_\nu \nonumber
\end{align}
\begin{center}
\rule{5.5cm}{0.01cm}
\end{center}
\begin{align}
{\cal O}_{\psi X1}&=g_s\bar u_L\sigma^{\mu\nu}G_{\mu\nu}u_R \qquad
{\cal O}_{\psi X2}=g_s\bar d_L\sigma^{\mu\nu}G_{\mu\nu}d_R \nonumber\\
{\cal O}_{\psi X3}&=
g\,\bar u_L\sigma^{\mu\nu} u_R\, (\partial_\mu W^3_\nu +ig W^+_\mu W^-_\nu)+
\sqrt{2}g\,\bar d_L\sigma^{\mu\nu} u_R\, (\partial_\mu W^-_\nu +ig W^-_\mu W^3_\nu)
\nonumber\\
{\cal O}_{\psi X4}&=
-g\,\bar d_L\sigma^{\mu\nu} d_R\, (\partial_\mu W^3_\nu +ig W^+_\mu W^-_\nu)+
\sqrt{2}g\,\bar u_L\sigma^{\mu\nu} d_R\, (\partial_\mu W^+_\nu -ig W^+_\mu W^3_\nu)
\nonumber\\
{\cal O}_{\psi X5}&=
-g\,\bar e_L\sigma^{\mu\nu} e_R\, (\partial_\mu W^3_\nu +ig W^+_\mu W^-_\nu)+
\sqrt{2}g\,\bar \nu_L\sigma^{\mu\nu} e_R\, (\partial_\mu W^+_\nu -igW^+_\mu W^3_\nu)
\nonumber\\
{\cal O}_{\psi X6}&=g' \bar u_L\sigma^{\mu\nu}B_{\mu\nu}u_R\qquad 
{\cal O}_{\psi X7} =g' \bar d_L\sigma^{\mu\nu}B_{\mu\nu}d_R \qquad
{\cal O}_{\psi X8} =g' \bar e_L\sigma^{\mu\nu}B_{\mu\nu}e_R\nonumber
\end{align}
\begin{center}
\rule{5.5cm}{0.01cm}
\end{center}
\begin{align}
{\cal O}_{LL6}&=
\bar q\gamma^\mu T_3 q\, \bar q\gamma_\mu T_3 q &
{\cal O}_{LL7}&=\bar q\gamma^\mu T_3 q\, \bar q\gamma_\mu q
\nonumber\\    
{\cal O}_{LL8}&=\bar q_\alpha\gamma^\mu T_3 q_\beta\, 
\bar q_\beta\gamma_\mu T_3 q_\alpha &
{\cal O}_{LL9}&=\bar q_\alpha\gamma^\mu  T_3 q_\beta\, 
\bar q_\beta\gamma_\mu  q_\alpha\nonumber\\
{\cal O}_{LL10}&=                                              
\bar q\gamma^\mu T_3 q\, \bar l\gamma_\mu T_3 l &
{\cal O}_{LL11}&=\bar q\gamma^\mu T_3 q\, \bar l\gamma_\mu l\nonumber\\
{\cal O}_{LL12}&=\bar q\gamma^\mu q\, \bar l\gamma_\mu T_3 l & 
{\cal O}_{LL13}&=\bar q\gamma^\mu  T_3 l\, 
\bar l\gamma_\mu  T_3 q\nonumber\\  
{\cal O}_{LL14}&=\bar q\gamma^\mu  T_3 l\, \bar l\gamma_\mu  q &
{\cal O}_{LL15}&=                                          
\bar l\gamma^\mu T_3 l\, \bar l\gamma_\mu T_3 l\nonumber\\ 
{\cal O}_{LL16}&=\bar l\gamma^\mu T_3 l\, \bar l\gamma_\mu l& \nonumber    
\end{align}
\begin{center}
\rule{5.5cm}{0.01cm}
\end{center}
\begin{align}
{\cal O}_{LR10}&=
\bar q\gamma^\mu T_3 q\, \bar u_R\gamma_\mu u_R &        
{\cal O}_{LR11}&=
\bar q\gamma^\mu T^A T_3 q\, \bar u_R\gamma_\mu T^A u_R\nonumber\\   
{\cal O}_{LR12}&=\bar q\gamma^\mu T_3 q\, \bar d_R\gamma_\mu d_R &               
{\cal O}_{LR13}&=
\bar q\gamma^\mu T^A T_3 q\, \bar d_R\gamma_\mu T^A d_R\nonumber\\
{\cal O}_{LR14}&=\bar u_R\gamma^\mu u_R\, \bar l\gamma_\mu T_3 l &           
{\cal O}_{LR15}&=\bar d_R\gamma^\mu d_R\, \bar l\gamma_\mu T_3 l\nonumber\\ 
{\cal O}_{LR16}&=\bar q\gamma^\mu T_3 q\, \bar e_R\gamma_\mu e_R &               
{\cal O}_{LR17}&=\bar l\gamma^\mu T_3 l\, \bar e_R\gamma_\mu e_R\nonumber\\ 
{\cal O}_{LR18}&=\bar q\gamma^\mu T_3 l\, \bar e_R\gamma_\mu d_R & \nonumber 
\end{align}
\begin{center}
\rule{5.5cm}{0.01cm}
\end{center}
\begin{align}
{\cal O}_{ST5}&=\bar u_L u_R\, \bar d_L d_R &
{\cal O}_{ST6}&=\bar d_L u_R\, \bar u_L d_R\nonumber\\                   
{\cal O}_{ST7}&=\bar u_L T^A u_R\, \bar d_L T^A d_R &
{\cal O}_{ST8}&=\bar d_L T^A u_R\, \bar u_L T^A d_R\nonumber\\                
{\cal O}_{ST9}&=\bar u_L u_R\, \bar e_L e_R &
{\cal O}_{ST10}&=\bar d_L u_R\, \bar \nu_L e_R\nonumber\\                 
{\cal O}_{ST11}&=\bar u_L\sigma^{\mu\nu} u_R\, \bar e_L\sigma_{\mu\nu} e_R &
{\cal O}_{ST12}&=
\bar d_L\sigma^{\mu\nu} u_R\, \bar \nu_L\sigma_{\mu\nu} e_R\nonumber
\end{align}
\begin{center}
\rule{5.5cm}{0.01cm}
\end{center}
\begin{align}                                   
{\cal O}_{FY1}&=\bar u_L u_R\, \bar u_L u_R &
{\cal O}_{FY2}&=\bar u_L T^A u_R\, \bar u_L T^A u_R\nonumber\\
{\cal O}_{FY3}&=\bar d_L d_R\, \bar d_L d_R &
{\cal O}_{FY4}&=\bar d_L T^A d_R\, \bar d_L T^A d_R\nonumber\\
{\cal O}_{FY5}&=\bar d_L d_R\, \bar u_R  u_L &
{\cal O}_{FY6}&=\bar d_L T^A d_R\, \bar u_R T^A u_L\nonumber\\
{\cal O}_{FY7}&=\bar d_L d_R\, \bar e_L e_R &
{\cal O}_{FY8}&=\bar d_L\sigma^{\mu\nu} d_R\, \bar e_L\sigma_{\mu\nu} e_R\nonumber\\
{\cal O}_{FY9}&=\bar e_L e_R\, \bar u_R u_L &
{\cal O}_{FY10}&=\bar e_L e_R\, \bar e_L e_R\nonumber\\
{\cal O}_{FY11}&=\bar e_L d_R\, \bar u_R \nu_L & \nonumber
\end{align}

\section*{Acknowledgements}

We thank Gino Isidori for comments on the manuscript and
useful discussions. We thank Witold Skiba for drawing our attention 
to ref. \cite{Grojean:2006nn}.
This work was performed in the context of the ERC Advanced Grant
project `FLAVOUR' (267104) and was supported in part by the 
DFG cluster of excellence `Origin and Structure of the Universe'.


\end{document}